\documentclass[aps,prb,floatfix,reprint]{revtex4-1}
\usepackage[T1]{fontenc}
\usepackage[latin9]{inputenc}
\usepackage[a4paper]{geometry}
\usepackage{xcolor}
\usepackage{verbatim}
\usepackage{textcomp}
\usepackage{amsmath}
\usepackage{amssymb}
\usepackage{stackrel}
\usepackage{graphicx}
\usepackage{wasysym}
\usepackage{hyperref}
\usepackage{xcolor,soul}

\begin{document}

\title{
Second-order adiabatic expansions of heat and charge currents with
nonequilibrium Green's functions.
}
\author{Sebasti\'an E. Deghi,$^1$ and Ra\'ul A. Bustos-Mar\'un$^{1,2}$}

\affiliation{$^1$Instituto de F\'{\i}sica Enrique Gaviola and Facultad de
Matem\'{a}tica, Astronom\'{\i}a, F\'{\i}sica y Computación, Universidad Nacional de
C\'{o}rdoba, Ciudad Universitaria, C\'{o}rdoba, 5000, Argentina\\
$^2$Facultad de Ciencias Qu\'{\i}micas, Universidad Nacional de C\'{o}rdoba, Ciudad
Universitaria, C\'{o}rdoba, 5000, Argentina}

\begin{abstract}
Due to technological needs, nanoscale heat management, energy conversion and quantum thermodynamics have become key areas of research, putting heat pumps and nanomotors center stage.
The treatment of these particular systems often requires the use of adiabatic expansions in terms of the frequency of the external driving or the velocity of some classical degree of freedom.
However, due to the difficulty of getting the expressions, most works have only explored first-order terms. Despite this, adiabatic expansions have allowed the study of intriguing phenomena such as adiabatic quantum pumps and motors, or electronic friction.
Here, we use nonequilibrium Green's functions, within a Schwinger-Keldysh approach, to develop second-order expressions for the energy, heat, and charge currents.
We illustrate, through two simple models, how the obtained formulas produce physically consistent results, and allow for the thermodynamic study of unexplored phenomena, such as second-order monoparametric pumping.\end{abstract}
\maketitle

\section{Introduction}

The conversion of heat into work (either mechanical or electrical)
has been at the center of technological and scientific interest since
the first studies of steam engines \citep{carnot1890}. In the past
decades, miniaturization has guided this interest toward micro- and
nanoscale heat-to-work conversion \citep{benenti2017,whitney2018,bustosmarun2019,zimbovskaya2020,aligia2020,pekola2021,alicki2021,eglinton2022,arrachea2023}.
This was not only due to the fundamental questions posed by the new
size scale \citep{esposito2015,acciai2024}, but also to practical
reasons such as the need for efficient heat management of nanodevices.
However, the nanoscale presents new challenges. On the one hand, the
regions of interest now involved a finite number of particles, as
opposed to classical thermodynamics. On the other hand, the explicit
treatment of quantum mechanical effects often becomes unavoidable.

Importantly, nanoscale quantum machines (devices with strong quantum
effects) require appropriate and efficient computational treatments.
This is so, since processes with vastly different time scales may
coexist there. Therefore, without any type of approximation, the calculations
must then be carried out with a temporal resolution given by the fastest
degrees of freedom (DOFs) but with total times determined by the slowest
DOFs. This clearly makes numerical calculations highly inefficient.
Even worse, sometimes contrasting theoretical approaches may be needed
for the different DOFs.

A successful strategy followed in the past consisted of treating some
slow DOFs classically (typically nuclear, mechanical, or external
ones). In contrast, the fast DOFs are treated fully quantum (typically
electrons or optical phonons). For nanoelectronic or nanoelectromechanical
devices, the time scale separation also drives the establishment of
steady-state currents which parametrically depend on the position
and velocity of the slow DOFs. This situation is amenable to some
kind of adiabatic approximation where the observable of interest is
expanded in terms of the frequency of the external driving or the
velocity of the classical DOFs. In this context, different theoretical
approaches have been used including non-equilibrium Green\textquoteright s
functions (NEGF) \citep{mozyrsky2006,pistolesi2008,zazunov2010,bode2011,deghi2021},
real-time diagrammatic approaches \citep{splettstoesser2006,leijnse2008,calvo2017,ribetto2021}
(or similar approaches based on the adiabatic expansions of the system's
reduced density matrix \citep{dundas2009,lin2019}), scattering formalism
\citep{moskalets2004,bennett2010,bode2011}, DFT-based calculations
\citep{lu2015}, and hierarchical equation of motion approaches \citep{erpenbeck2018,rudge2023}.

Within the context of quantum transport, adiabatic approximations
up to the first order have allowed the treatment of fascinating phenomena
such as adiabatic quantum pumping \citep{brouwer1998}, adiabatic
quantum motors\citep{bustosmarun2013}, negative friction coefficients
\citep{bode2011,preston2023}, reciprocity breaking \citep{wachtler2023,mehring2024},
hysteresis \citep{kurnosov2022,mehring2024}, current-noise induced
by thermal oscillations \citep{ribetto2023}, etc. Regardless of the
notable advances made, one may wonder about the unexplored phenomena
that may await beyond first-order adiabatic treatments. In this sense,
Kershaw et al. \citep{kershaw2017,kershaw2019} pioneered the second-order
treatment of electric currents within a NEGF approach. However, a
thermodynamically consistent second-order treatment of quantum machines
necessarily requires addressing heat currents. Moreover, explicit
expressions of the observables are desirable for numerical calculations,
instead of implicit expressions that need to be developed for the
cases of interest. In the present manuscript, we present the complete
and explicit expressions, up to the second order of the adiabatic
expansions of heat and charge currents. Our results are based on NEGFs
within a Schwinger-Keldysh approach. The obtained formulas are general
and restricted only by the assumption of time-independent self-energies.\textcolor{red}{}\footnote{It is important to highlight that this quite common assumption is
not a limitation for a wide class of systems. For example, in a tight-binding
approach, one can redefine the local system by adding the sites of
the leads with time dependence \citep{cattena2014}.}

This manuscript is organized as follows. Sec. \ref{sec: General Theory.}
presents the general theory based on the NEGF formalism. Having developed
the theoretical tools, in Sec. \ref{sec: Observables.}, we provide
the second-order adiabatic expansion for charge, energy, and heat
currents. Sec. \ref{sec: Models.} contains two different models illustrating
how the corrections for each order work and the usefulness of the
expressions for thermodynamic analysis. In particular, the last example
treats second-order quantum pumping, a phenomenon not previously described
up to our knowledge. The examples also include a numerical verification
of our formulas through the first law of thermodynamics. Finally,
Sec. \ref{sec: Conclusion.} comprises a summary and a brief discussion.
In an effort not to overload the readers, all the nonessential comments
and mathematical details lay in the Appendices from \ref{sec: Appendix A - Wigner Transform of Dyson Equation.}
to \ref{sec: Appendix I -Current Induced Forces.}.

\section{General theory\label{sec: General Theory.}}

\subsection{Generic Hamiltonian}

To define the different parts of the system and their associated creation
(annihilation) operators, we introduce the total Hamiltonian $H$
that describes the \textit{total system}. It consists of a core region
$S$ connected to several macroscopic reservoirs $\alpha$ (the leads).
The core region, which typically has nanometric dimensions, will be
referred to as the \textit{local system}. The total Hamiltonian takes
the form
\begin{align*}
H & =\underset{\alpha}{\sum}H_{\alpha}+H_{S}+\underset{\alpha}{\sum}V_{\alpha},
\end{align*}
where $H_{S}$, in general, corresponds to a time-dependent local
system, $H_{\alpha}$ describes the time-independent $\alpha$-lead,
and the term $V_{\alpha}$ contains the time-independent interaction
between the local system and the $\alpha$-lead. The Hamiltonian of
the local system depends parametrically on time through a multidimensional
vector $\overrightarrow{X}$, which describes the mechanical degrees
of freedom. The local system's Hamiltonian assumes the following form
when the second quantization is applied
\begin{align}
H_{S} & =\underset{l,s}{\sum}h_{l,s}\left(\overrightarrow{X}\right)d_{l}^{\dagger}d_{s}.\label{eq: Local System Hamiltonian - Definition.}
\end{align}
In the notation we have used, the operators $d_{l}^{\dagger}$ and
$d_{l}$ create or annihilate, respectively, an electron within the
local system. The Hamiltonians of the leads can be written as
\begin{align}
H_{\alpha} & =\underset{k}{\sum}\epsilon_{\alpha k}c_{\alpha k}^{\dagger}c_{\alpha k},\label{eq: Alpha Lead Hamiltonian - Definition.}
\end{align}
where $\epsilon_{\alpha k}$ is the energy of the $\alpha$-lead in
the state $k$, $c_{\alpha k}^{\dagger}$ creates an electron in the
$\alpha$-lead with the state $k$, whereas the $c_{\alpha k}$ annihilates
it. For simplicity, the index $k$ also includes the electron's spin.
Finally, the tunneling interaction describes the coupling between
the local system and the $\alpha$-leads
\begin{align*}
V_{\alpha} & =\underset{k,l}{\sum}\left(t_{\alpha k,l}c_{\alpha k}^{\dagger}d_{l}+t_{\alpha k,l}^{*}d_{l}^{\dagger}c_{\alpha k}\right),
\end{align*}
where $t_{\alpha k,l}$ are the tunneling amplitudes between the leads
and the local system. 

\subsection{Dynamics of non-equilibrium open quantum systems}

Within the context of the Schwinger-Keldysh approach for the NEGF
formalism \citep{jauho1994,maciejko2007,haug2008book,spicka2014,odashima2017}
we define the elements of the retarded and advanced Green\textquoteright s
functions, $\mathcal{G}^{R}$ and $\mathcal{G}^{A}$
respectively. For a system that evolves, not necessarily in an equilibrium
process, from $t'$ to $t$ times, they
are
\begin{align}
\mathcal{G}_{l,s}^{R}\left(t,t'\right) & =-\frac{i}{\hbar}\Theta\left(t-t'\right)\left\langle \left\{ d_{s}\left(t\right),d_{l}^{\dagger}\left(t'\right)\right\} \right\rangle ,\label{eq: Retarded Green's Funtion - Formal Definition.}
\end{align}
and
\begin{align}
\mathcal{G}_{s,l}^{A}\left(t',t\right) & =\left[\mathcal{G}_{l,s}^{R}\left(t,t'\right)\right]^{*},\label{eq: Advanced Green's Funtion - Formal Definition.}
\end{align}
where $\left\{ \bullet,\bullet\right\} $ is the anticommutator and
$\left\langle \bullet\right\rangle $ is the quantum expectation value.

For the states of the local system, the lesser Green\textquoteright s
function elements are given by
\begin{align}
\mathcal{G}_{l,s}^{<}\left(t,t'\right) & =\frac{i}{\hbar}\left\langle d_{l}^{\dagger}\left(t'\right)d_{s}\left(t\right)\right\rangle ,\label{eq: Lesser Green's Funtion - Formal Definition - System betwen System.}
\end{align}
whereas for the states that propagate between the local system and
leads, the lesser Green\textquoteright s function elements are
\begin{align}
\mathcal{G}_{\alpha k,l}^{<}\left(t,t'\right) & =\frac{i}{\hbar}\left\langle c_{\alpha k}^{\dagger}\left(t'\right)d_{l}\left(t\right)\right\rangle .\label{eq: Lesser Green's Funtion - Formal Definition - System betwen Lead.}
\end{align}

Furthermore, the retarded and advanced self-energies are assumed to
be stationary.\footnote{For our purposes, stationary implies that the self-energies only depend
on the time difference $t-t'$, which means that both $V_{\alpha}$ and $H_{\alpha}$ are independent of the mechanical
degrees of freedom.} The elements of the retarded self-energy for the $\alpha$-lead take
the form
\begin{align}
\Sigma_{\alpha,l,s}^{R}\left(t,t'\right) & =\underset{k}{\sum}t_{l,\alpha k}g_{\alpha k}^{R}\left(t,t'\right)t_{\alpha k,s},\label{eq: Alpha Lead Retarded Self Energy - Time Domain.}
\end{align}
where
\begin{align}
g_{\alpha k}^{R}\left(t,t'\right) & =-\frac{i}{\hbar}\Theta\left(t-t'\right)e^{-\frac{i}{\hbar}\epsilon_{\alpha k}\left(t-t'\right)}.\label{eq: Retarded Self Energies Kernels}
\end{align}
Hence, the elements of total retarded self-energy comprise the sum
of all leads
\begin{align}
\Sigma_{l,s}^{R}\left(t,t'\right) & =\underset{\alpha}{\sum}\Sigma_{\alpha,l,s}^{R}\left(t,t'\right).\label{eq: Total Retarded Self Energy - Time Domain.}
\end{align}
To obtain the advanced self-energy, we calculate the adjoint of the
retarded one
\begin{align}
\Sigma_{\alpha,s,l}^{A}\left(t',t\right) & =\left[\Sigma_{\alpha,l,s}^{R}\left(t,t'\right)\right]^{*}.\label{eq: Alpha Lead Advanced Self Energy - Time Domain.}
\end{align}
Likewise, the elements of the lesser self-energy of the $\alpha$-lead
are given by
\begin{align}
\Sigma_{\alpha,l,s}^{<}\left(t,t'\right) & =\underset{k}{\sum}t_{l,\alpha k}g_{\alpha k}^{<}\left(t,t'\right)t_{\alpha k,s},\label{eq: Alpha Lead Lesser Self Energy - Time Domain.}
\end{align}
where
\begin{align}
g_{\alpha k}^{<}\left(t,t'\right) & =\frac{2\pi i}{\hbar}f_{\alpha}\left(\epsilon_{\alpha k}\right)e^{-\frac{i}{\hbar}\epsilon_{\alpha k}\left(t-t'\right)}.\label{eq: Lesser Self Energies Kernels}
\end{align}
Here, $f_{\alpha}$ is the $\alpha$-lead's Fermi-Dirac distribution.
To conclude, the elements of the total lesser self-energy are
\begin{align*}
\Sigma_{l,s}^{<}\left(t,t'\right) & =\underset{\alpha}{\sum}\Sigma_{\alpha,l,s}^{<}\left(t,t'\right).
\end{align*}

Given the essential operators, the next step is to delve into the
quantum dynamics, which in Schwinger-Keldysh formalism is expressed
by the integrodifferential Dyson equation in terms of the retarded
Green\textquoteright s function 
\begin{align}
-i\hbar\partial_{t'}\mathcal{G}^{R}\left(t,t'\right) & =\delta\left(t-t'\right)+\mathcal{G}^{R}\left(t,t'\right)H_{S}\left(t'\right)\label{eq: Differential Dyson Equation for Green's Funtion.}\\
 & +\int\mathcal{G}^{R}\left(t,t_{1}\right)\Sigma^{R}\left(t_{1},t'\right)dt_{1}.\nonumber 
\end{align}
Here, the Hamiltonian of the local system, retarded Green\textquoteright s
function and self-energy are given by Eqs. (\ref{eq: Local System Hamiltonian - Definition.}),
(\ref{eq: Retarded Green's Funtion - Formal Definition.}) and (\ref{eq: Total Retarded Self Energy - Time Domain.}),
respectively.

At this point, we can apply many techniques to solve Eq. (\ref{eq: Differential Dyson Equation for Green's Funtion.}).
One way is to use numerical methods \citep{kloss2021}. Another one
is to develop an adiabatic expansion of the observables. To carry
out the latter technique, we must decompose the defined operators
and Dyson\textquoteright s equation into two-time scales, which is
done through the use of a Wigner transform.

\subsection{Adiabatic expansion for retarded Green\textquoteright s function}

The idea is to separate the fast microscopic dynamics, followed by
the electrons, from the slow macroscopic changes, determined by the
classical mechanical degrees of freedom. Once the time scales have
been distinguished, a gradient expansion will be carried out on the
slow variables. First of all, we have to define the following transform
of time variables
\begin{align}
T & =\frac{t+t'}{2},\label{eq: Wigner Transform - Slow Coordinate.}\\
\tau & =t-t'.\label{eq: Wigner Transform - Fast Coordinate.}
\end{align}
The time scale $T$ is called the slow variable, while $\tau$ is
the fast variable. The Wigner transform for the retarded Green\textquoteright s
function is defined as
\begin{align*}
\mathcal{\widetilde{G}}^{R}\left(T,\varepsilon\right) & =\int\mathcal{G}^{R}\left(t,t'\right)e^{\frac{i}{\hbar}\varepsilon\tau}d\tau,
\end{align*}
whereas its inverse transformation is
\begin{align*}
\mathcal{G}^{R}\left(t,t'\right) & =\frac{1}{2\pi\hbar}\int\mathcal{\widetilde{G}}^{R}\left(T,\varepsilon\right)e^{-\frac{i}{\hbar}\varepsilon\tau}d\varepsilon.
\end{align*}
For the elements of the retarded self-energy described in Eq. (\ref{eq: Alpha Lead Retarded Self Energy - Time Domain.}),
$\Sigma^{R}\left(t,t'\right)\equiv\Sigma^{R}\left(\tau\right)$, the
Wigner transform becomes a Fourier transform, where

\begin{eqnarray*}
\Sigma^{R}\left(\varepsilon\right) & = & \int\Sigma^{R}\left(\tau\right)e^{\frac{i}{\hbar}\varepsilon\tau}d\tau,
\end{eqnarray*}
and

\begin{align}
\Sigma_{\alpha,l,s}^{R}\left(\varepsilon\right) & =\Delta_{\alpha,l,s}\left(\varepsilon\right)-i\Gamma_{\alpha,l,s}\left(\varepsilon\right).\label{eq: Wigner Transform - Retarded Self Energy.}
\end{align}
Here, the level-width functions $\Gamma_{\alpha,l,s}$ are given by\footnote{The definitions of the level-width functions may differ from that
of other authors \citep{haug2008book,kershaw2019}. Here, we followed
the convention used in\textcolor{blue}{{} \citep{bode2011,cattena2014,deghi2021}}.}
\begin{align*}
\Gamma_{\alpha,l,s}\left(\varepsilon\right) & =\pi\underset{k}{\sum}\delta\left(\varepsilon-\epsilon_{\alpha k}\right)t_{l,\alpha k}t_{\alpha k,s}.
\end{align*}
The level-shift functions $\Delta_{\alpha,l,s}$ can be calculated
from $\Gamma_{\alpha,l,s}$ using the Kramers-Kroning relation \citep{haug2008book}.
After applying the adjoint operator to Eq. (\ref{eq: Wigner Transform - Retarded Self Energy.})
we get the advanced self-energy $\Sigma_{\alpha,l,s}^{A}$. For the
elements of the lesser self-energy given in Eq. (\ref{eq: Alpha Lead Lesser Self Energy - Time Domain.}),
their Wigner transform give
\begin{align}
\Sigma_{\alpha,l,s}^{<}\left(\varepsilon\right) & =2if_{\alpha}\left(\varepsilon\right)\Gamma_{\alpha,l,s}\left(\varepsilon\right).\label{eq: Wigner Transform - Lesser Self Energy.}
\end{align}
Given the operators in the energy domain, we must apply the Moyal
product to Eq. (\ref{eq: Differential Dyson Equation for Green's Funtion.}),
yielding (see Appendix \ref{sec: Appendix A - Wigner Transform of Dyson Equation.})
\begin{align}
1 & =-\frac{1}{2}i\hbar\partial_{T}\mathcal{\widetilde{G}}^{R}\left(T,\varepsilon\right)+\varepsilon\mathcal{\widetilde{G}}^{R}\left(T,\varepsilon\right)\nonumber \\
 & -\mathcal{\widetilde{G}}^{R}\left(T,\varepsilon\right)e^{\frac{i\hbar}{2}\left(\overleftarrow{\partial}_{\varepsilon}\overrightarrow{\partial}_{T}-\overleftarrow{\partial}_{T}\overrightarrow{\partial}_{\varepsilon}\right)}\Sigma^{R}\left(\varepsilon\right)\label{eq: Retarded Green Funtion - Moyal Product.}\\
 & -\mathcal{\widetilde{G}}^{R}\left(T,\varepsilon\right)e^{\frac{i\hbar}{2}\left(\overleftarrow{\partial}_{\varepsilon}\overrightarrow{\partial}_{T}-\overleftarrow{\partial}_{T}\overrightarrow{\partial}_{\varepsilon}\right)}\hat{H}_{S}\left(T\right),\nonumber 
\end{align}
The above expression implies that the unknown retarded green's function
$\mathcal{\widetilde{G}}^{R}$ can be expressed as an infinite sum
over its derivatives, which are also unknown in principle. To solve
this, we have to implement an iterative approach (see Appendix \ref{sec: Apendix B - Adiabatic Expansion for Retarded Green's Functions.})
which starts at zero-order by approximating $\mathcal{\widetilde{G}}^{R}$
by the adiabatic Green's function $G^{R}$, defined as
\begin{align}
G^{R} & =\left[\varepsilon I-H_{S}\left(T\right)-\Sigma^{R}\left(\varepsilon\right)\right]^{-1}.\label{eq: Adiabatic Retarded Green Function}
\end{align}
The above adiabatic (or frozen) Green's function $G^{R}$ is formally
equal to $\mathcal{\widetilde{G}}^{R}\left(T,\varepsilon\right)$
in the limit of infinitely slow mechanical degrees of freedom, which
constitute a form of the Born-Oppenheimer approximation.

By the chain rule, the slow-time $T$ derivatives of the Hamiltonian
of the local system, up to the second order, are
\begin{align}
\partial_{T}H_{S} & =\stackrel[\nu=1]{M}{\sum}\varLambda_{\nu}\dot{X}_{\nu},\label{eq: First Slow Time Derivative - Local Hamiltonian.}\\
\partial_{T}^{2}H_{S} & =\stackrel[\mu,\nu=1]{M}{\sum}\varLambda_{\mu\nu}\dot{X}_{\mu}\dot{X}_{\nu}+\stackrel[\nu=1]{M}{\sum}\varLambda_{\nu}\ddot{X}_{\nu}.\label{eq: Second Slow Time Derivative - Local Hamiltonian.}
\end{align}
Here, $\dot{X}_{\nu}$ and $\ddot{X}_{\nu}$ are the time derivatives
of the mechanical degree of freedom $\nu$, $\partial_{T}X_{\nu}$
and $\partial_{T}^{2}X_{\nu}$ respectively. Additionally, the elements
of the matrices $\varLambda_{\nu}$ and $\varLambda_{\mu\nu}$ are
given by
\begin{align*}
\left[\varLambda_{\nu}\right]_{l,s} & =\frac{\partial h_{l,s}}{\partial X_{\nu}}, & \left[\varLambda_{\mu\nu}\right]_{l,s} & =\frac{\partial^{2}h_{l,s}}{\partial X_{\mu}\partial X_{\nu}}.
\end{align*}
Then, applying the iterative method twice to second-order terms, the
resulting retarded Green's function takes the form
\begin{align}
\mathcal{\widetilde{G}}^{R} & \simeq G^{R}\left(T,\varepsilon\right)+\frac{i\hbar}{2}\stackrel[\nu=1]{M}{\sum}\Omega_{1,\nu}^{R}\left(T,\varepsilon\right)\dot{X}_{\nu}\label{eq: Adiabatic Expansion of Retarded Green's Function.}\\
 & +\left(\frac{i\hbar}{2}\right)^{2}\left\{ \stackrel[\mu,\nu=1]{M}{\sum}\Omega_{11,\mu\nu}^{R}\dot{X}_{\mu}\dot{X}_{\nu}+\stackrel[\nu=1]{M}{\sum}\Omega_{2,\nu}^{R}\ddot{X}_{\nu}\right\} .\nonumber 
\end{align}
To make the notation more compact, we have defined the following operators
\begin{align*}
\Omega_{1,\nu}^{R} & =\Xi_{1}\left[G^{R},\varLambda_{\nu},G^{R}\right],\\
\Omega_{11,\mu\nu}^{R} & =\Xi_{11}\left[G^{R},K_{\mu\nu}^{R},G^{R}\right],\\
\Omega_{2,\nu}^{R} & =\frac{1}{2}\Xi_{2}\left[G^{R},\varLambda_{\nu},G^{R}\right].
\end{align*}
where we have also introduced the definitions
\begin{align*}
K_{\mu\nu}^{R} & =\varLambda_{\mu}G^{R}\left(T,\varepsilon\right)\varLambda_{\nu}+\frac{1}{2}\varLambda_{\mu\nu},
\end{align*}
and
\begin{align*}
\Xi_{1}\left[A,B,C\right] & =\left(\partial_{\varepsilon}A\right)BC-AB\left(\partial_{\varepsilon}C\right),\\
\Xi_{2}\left[A,B,C\right] & =\left(\partial_{\varepsilon}^{2}A\right)BC-2\left(\partial_{\varepsilon}A\right)B\left(\partial_{\varepsilon}C\right)\\
 & +AB\left(\partial_{\varepsilon}^{2}C\right),\\
\Xi_{11}\left[A,B,C\right] & =\Xi_{2}\left[A,B,C\right]-A\left(\partial_{\varepsilon}^{2}B\right)C.
\end{align*}

Take into account that the order to which the Planck's constant is
elevated is consistent with the order of the adiabatic expansion.
In this sense, it can be used as a \textquotedbl book-keeping\textquotedbl{}
parameter to keep track of the expansion order for a given term.

Finally, to calculate the advanced Green function, we use the relation
\begin{align*}
\mathcal{\widetilde{G}}^{A} & =\left[\mathcal{\widetilde{G}}^{R}\left(T,\varepsilon\right)\right]^{\dagger}.
\end{align*}

\subsection{Adiabatic expansion for lesser Green\textquoteright s function}

To find the Wigner\textquoteright s transform of the lesser Green\textquoteright s
functions, Eqs. (\ref{eq: Lesser Green's Funtion - Formal Definition - System betwen System.})
and (\ref{eq: Lesser Green's Funtion - Formal Definition - System betwen Lead.}),
we start with the relation \textcolor{blue}{\citep{maciejko2007,haug2008book}}
\begin{align*}
\mathcal{G}^{<}\left(t,t'\right) & =\iint\mathcal{G}^{R}\left(t,t_{1}\right)\Sigma^{<}\left(t_{1},t_{2}\right)\mathcal{G}^{A}\left(t_{2},t'\right)dt_{1}dt_{2}.
\end{align*}
After applying the Wigner transform to both sides of the above formula
and, afterwards, the Moyal product, we arrive at the expression
\begin{align}
\mathcal{\widetilde{G}}^{<} & =\mathcal{\widetilde{G}}^{R}\left(T,\varepsilon\right)e^{\frac{i\hbar}{2}\left(\overleftarrow{\partial}_{\varepsilon}\overrightarrow{\partial}_{T}-\overleftarrow{\partial}_{T}\overrightarrow{\partial}_{\varepsilon}\right)}\Sigma^{<}\left(\varepsilon\right)\times\nonumber \\
 & e^{\frac{i\hbar}{2}\left(\overleftarrow{\partial}_{\varepsilon}\overrightarrow{\partial}_{T}-\overleftarrow{\partial}_{T}\overrightarrow{\partial}_{\varepsilon}\right)}\mathcal{\widetilde{G}}^{A}\left(T,\varepsilon\right).\label{eq: Lesser Green Funtion - Moyal Product.}
\end{align}
As we discussed for the retarded Green\textquoteright s function,
$\mathcal{\widetilde{G}}^{<}$ takes the form of a set of infinite
sums (see Appendix \ref{sec: Appendix C - Adiabatic Expansion for Lesser Green's Functions.}),
where the elements of $\Sigma^{<}$ are given in Eq. (\ref{eq: Wigner Transform - Lesser Self Energy.}).
To solve this, we first insert in Eq. (\ref{eq: Lesser Green Funtion - Moyal Product.})
the second-order expansion of the operators $\mathcal{\widetilde{G}}^{R}$
and $\mathcal{\widetilde{G}}^{A}$ derived before, and then we truncate
the series at second order. This yields
\begin{align}
\mathcal{\widetilde{G}}^{<} & \simeq G^{<}\left(T,\varepsilon\right)+\frac{i\hbar}{2}\stackrel[\nu=1]{M}{\sum}\Omega_{1,\nu}^{<}\left(T,\varepsilon\right)\dot{X}_{\nu}\label{eq: Adiabatic Expansion of Lesser Green's Function.}\\
 & +\left(\frac{i\hbar}{2}\right)^{2}\left\{ \stackrel[\mu,\nu=1]{M}{\sum}\Omega_{11,\mu\nu}^{<}\dot{X}_{\mu}\dot{X}_{\nu}+\stackrel[\nu=1]{M}{\sum}\Omega_{2,\nu}^{<}\ddot{X}_{\nu}\right\} ,\nonumber 
\end{align}
Where the adiabatic lesser Green function is
\begin{align*}
G^{<} & =G^{R}\left(T,\varepsilon\right)\Sigma^{<}\left(\varepsilon\right)G^{A}\left(T,\varepsilon\right),
\end{align*}
and the remaining operators are defined by
\begin{align*}
\Omega_{1,\nu}^{<} & =\Xi_{1}\left[G^{R},\varLambda_{\nu},G^{<}\right]+\Xi_{1}\left[G^{<},\varLambda_{\nu},G^{A}\right],\\
\Omega_{11,\mu\nu}^{<} & =\Xi_{11}\left[G^{R},K_{\mu\nu}^{R},G^{<}\right]+\Xi_{11}\left[G^{<},K_{\mu\nu}^{A},G^{A}\right]\\
 & +\Xi_{11}\left[G^{R},\varLambda_{\mu}G^{<}\varLambda_{\nu},G^{A}\right],\\
\Omega_{2,\nu}^{<} & =\frac{1}{2}\left\{ \Xi_{2}\left[G^{<},\varLambda_{\nu},G^{A}\right]+\Xi_{2}\left[G^{R},\varLambda_{\nu},G^{<}\right]\right\} .
\end{align*}
In the above expressions we have introduced the following term to
compact the notation
\begin{align*}
K_{\mu\nu}^{A} & =\varLambda_{\mu}G^{A}\left(T,\varepsilon\right)\varLambda_{\nu}+\frac{1}{2}\varLambda_{\mu\nu}.
\end{align*}

\section{Observables\label{sec: Observables.}}

We have already developed explicit formulas for evaluating up-to-second-order
corrections to the adiabatic Green's functions. In the following section,
we will apply these results to obtain close expressions for the second-order
adiabatic corrections of three observables of interest within quantum
transport: charge, energy, and heat currents.

\subsection{Charge current\label{subsec:Charge-current}}

For an $\alpha$-lead, the average charge current through it can be
written as the mean value of the time derivative of the number operator
$N_{\alpha}$
\begin{align}
I_{\alpha} & =-e\left\langle \dot{N}_{\alpha}\left(t\right)\right\rangle ,\label{eq: Charge Current =002013 Definition.}
\end{align}
where $N_{\alpha}=\underset{k}{\sum}c_{\alpha k}^{\dagger}c_{\alpha k},$ and
$e$ is the modulus of the electron charge. Then, applying Ehrenfest\textquoteright s
theorem, the commutation relations for fermions, and using Eqs. (\ref{eq: Lesser Green's Funtion - Formal Definition - System betwen System.})
and (\ref{eq: Lesser Green's Funtion - Formal Definition - System betwen Lead.})
in Eq. (\ref{eq: Charge Current =002013 Definition.}), we get a formula
for the charge current in terms of the lesser Green's function
\begin{align}
I_{\alpha} & =2e\textrm{Re}\left\{ \underset{k,l}{\sum}t_{\alpha k,l}\mathcal{G}_{l,\alpha k}^{<}\left(t,t\right)\right\} .\label{eq: Carge Current =002013 First Expression - Time Domain.}
\end{align}

To evaluate the time-dependent current $I_{\alpha}$, we must proceed
in accord with the Langreth\textquoteright s rules \citep{vanLeeuwen2006,haug2008book},
yielding
\begin{align}
I_{\alpha} & =2e\int\textrm{Re}\left\{ \textrm{tr}\left[\mathcal{G}^{R}\left(t,t'\right)\varSigma_{\alpha}^{<}\left(t',t\right)\right.\right.\nonumber \\
 & \left.\left.+\mathcal{G}^{<}\left(t,t'\right)\varSigma_{\alpha}^{A}\left(t',t\right)\right]\right\} dt'.\label{eq: Charge Current =002013 Second Expression - Time Domain.}
\end{align}

The next step is to insert the Wigner transform of the Green's functions
into Eq. (\ref{eq: Charge Current =002013 Second Expression - Time Domain.})
and make a gradient expansion, the resulting expression reads (see
Appendix \ref{sec: Appendix D - Adiabatic expansion for particle's current.})
\begin{align}
I_{\alpha} & =\frac{2e}{h}\stackrel[N=0]{\infty}{\sum}\frac{\left(-1\right)^{N}}{N!}\int\textrm{Re}\left\{ \textrm{tr}\left\{ C_{\alpha,N}^{I}\right\} \right\} d\varepsilon,\label{eq: Charge Current =002013 Second Expression - Energy Domain.}
\end{align}
where
\begin{align*}
C_{\alpha,N}^{I} & =\left(\frac{i\hbar}{2}\right)^{N}\left(\partial_{T}^{N}\mathcal{\widetilde{G}}^{R}\partial_{\varepsilon}^{N}\varSigma_{\alpha}^{<}+\partial_{T}^{N}\mathcal{\widetilde{G}}^{<}\partial_{\varepsilon}^{N}\varSigma_{\alpha}^{A}\right).
\end{align*}

It is important to highlight that, although this expression is exact,
its explicit evaluation necessarily requires approximations. For this
purpose, we will use the up-to-second-order adiabatic expansion
of the Green's functions found previously. After some algebra, one
can find the up-to-second-order adiabatic expansion of the charge
current, giving
\begin{align}
I_{\alpha} & \simeq I_{\alpha}^{\left(0\right)}+I_{\alpha}^{\left(1\right)}+I_{\alpha}^{\left(1,1\right)}+I_{\alpha}^{\left(2\right)}.\label{eq: Charge Current =002013 Second Order Adiabatic Expansion.}
\end{align}
The first two terms of the above expression are well known. The former,
$I_{\alpha}^{\left(0\right)}$, is equivalent to Landauer's formula,
while the second one, $I_{\alpha}^{\left(1\right)}$, is the quantum
pumping contribution to the charge current \citep{brouwer1998,bode2011}
\begin{align}
I_{\alpha}^{\left(0\right)} & =-e\mathcal{N}_{\alpha}^{\left(0\right)},\label{eq: Charge Current =002013 Cero Order Adiabatic Term.}\\
I_{\alpha}^{\left(1\right)} & =-e\underset{\nu}{\sum}\mathcal{N}_{\alpha,\nu}^{\left(1\right)}\dot{X}_{\nu}.\label{eq: Charge Current =002013 First Order Adiabatic Term.}
\end{align}
Here, we have introduced the terms 
\begin{align*}
\mathcal{N}_{\alpha}^{\left(0\right)} & =\int\mathcal{K}_{\alpha}^{\left(0\right)}\left(T,\varepsilon\right)d\varepsilon, & \mathcal{N}_{\alpha,\nu}^{\left(1\right)} & =\int\mathcal{K}_{\alpha,\nu}^{\left(1\right)}\left(T,\varepsilon\right)d\varepsilon.
\end{align*}
We acquire the preceding definition for practical purposes, which
will be useful later, being the kernel\textquoteright s integrals
defined as
\begin{align*}
\mathcal{K}_{\alpha}^{\left(0\right)} & =-\frac{2}{h}\textrm{Re}\left\{ \textrm{tr}\left(\Phi_{\alpha}^{\left(0\right)}\right)\right\} ,\\
\mathcal{K}_{\alpha,\nu}^{\left(1\right)} & =\frac{1}{\pi}\textrm{Im}\left\{ \textrm{tr}\left(\partial_{\varepsilon}G^{R}\varLambda_{\nu}\Phi_{\alpha}^{\left(0\right)}+\partial_{\varepsilon}G^{<}\varLambda_{\nu}\Phi_{\alpha}^{\left(1\right)}\right)\right\} ,
\end{align*}
where
\begin{align*}
\Phi_{\alpha}^{\left(0\right)} & =G^{r}\varSigma_{\alpha}^{<}\left(\varepsilon\right)+G^{<}\varSigma_{\alpha}^{A}\left(\varepsilon\right), & \Phi_{\alpha}^{\left(1\right)} & =G^{A}\varSigma_{\alpha}^{A}\left(\varepsilon\right).
\end{align*}

The remaining terms in Eq. (\ref{eq: Charge Current =002013 Second Order Adiabatic Expansion.})
constitute the second-order adiabatic terms of the charge current.
These formulas can be written as
\begin{align}
I_{\alpha}^{\left(2\right)} & =-e\underset{\nu}{\sum}\mathcal{N}_{\alpha,\nu}^{\left(2\right)}\ddot{X}_{\nu},\label{eq: Charge Current =002013 Second Order Adiabatic Term =002013 Second Derivative.}\\
I_{\alpha}^{\left(1,1\right)} & =-e\underset{\mu,\nu}{\sum}\mathcal{N}_{\alpha,\mu\nu}^{\left(1,1\right)}\dot{X}_{\mu}\dot{X}_{\nu}.\label{eq: Charge Current =002013 Second Order Adiabatic Term =002013 First Cross Derivatives.}
\end{align}
The current derivatives $\mathcal{N}_{\alpha,\nu}^{\left(2\right)}=-\frac{1}{e}\frac{\partial I_{\alpha}}{\partial\ddot{X}_{\nu}}$
and $\mathcal{N}_{\alpha,\mu\nu}^{\left(1,1\right)}=-\frac{1}{e}\frac{\partial^{2}I_{\alpha}}{\partial\dot{X}_{\mu}\partial\dot{X}_{\nu}}$
are
\begin{align*}
\mathcal{N}_{\alpha,\nu}^{\left(2\right)} & =\int\mathcal{K}_{\alpha,\nu}^{\left(2\right)}\left(T,\varepsilon\right)d\varepsilon, & \mathcal{N}_{\alpha,\mu\nu}^{\left(1,1\right)} & =\int\mathcal{K}_{\alpha,\mu\nu}^{\left(1,1\right)}\left(T,\varepsilon\right)d\varepsilon,
\end{align*}
The above kernels of integrals are
\begin{align*}
\mathcal{K}_{\alpha,\nu}^{\left(2\right)} & =\frac{\hbar}{2\pi}\textrm{Re}\left\{ \textrm{tr}\left\{ \partial_{\varepsilon}^{2}G^{R}\varLambda_{\nu}\Phi_{\alpha}^{\left(0\right)}+\partial_{\varepsilon}^{2}G^{<}\varLambda_{\nu}\Phi_{\alpha}^{\left(1\right)}\right\} \right\} ,\\
\mathcal{K}_{\alpha,\mu\nu}^{\left(1,1\right)} & =\frac{\hbar}{\pi}\textrm{Re}\left\{ \textrm{tr}\left\{ \Phi_{\mu\nu}^{\left(2\right)}\Phi_{\alpha}^{\left(0\right)}+\Phi_{\mu\nu}^{\left(3\right)}\Phi_{\alpha}^{\left(1\right)}\right\} \right\} ,
\end{align*}
where
\begin{align*}
\Phi_{\mu\nu}^{\left(2\right)} & =\partial_{\varepsilon}\left(\partial_{\varepsilon}G^{R}K_{\mu\nu}^{R}\right),\\
\Phi_{\mu\nu}^{\left(3\right)} & =\partial_{\varepsilon}\left(\partial_{\varepsilon}G^{R}\varLambda_{\mu}G^{<}\varLambda_{\nu}+\partial_{\varepsilon}G^{<}K_{\mu\nu}^{A}\right).
\end{align*}

\subsection{Energy current\label{subsec:Energy-current}}

Analogous to the charge current, the energy current flowing in the
$\alpha$-lead $\dot{E}_{\alpha}$ is defined as the mean value of
the time derivative of the Hamiltonian of the $\alpha$-lead $H_{\alpha}$,
\begin{align}
\dot{E}_{\alpha} & =\left\langle \dot{H}_{\alpha}\left(t\right)\right\rangle .\label{eq: Energy Current =002013 Definition.}
\end{align}
As we proceeded in the previous section for the charge current, we
will apply Ehrenfest's theorem, the commutation relations for fermions,
and Eqs. (\ref{eq: Lesser Green's Funtion - Formal Definition - System betwen System.})
and (\ref{eq: Lesser Green's Funtion - Formal Definition - System betwen Lead.})
into Eq. (\ref{eq: Energy Current =002013 Definition.}). This gives
\begin{align}
\dot{E}_{\alpha} & =-2\textrm{Re}\left\{ \underset{k,l}{\sum}\epsilon_{\alpha k}t_{\alpha k,l}\mathcal{G}_{l,\alpha k}^{<}\left(t,t\right)\right\} .\label{eq: Energy Current =002013 First Expression - Time Domain.}
\end{align}
Note that the essential difference between Eqs. (\ref{eq: Carge Current =002013 First Expression - Time Domain.})
and (\ref{eq: Energy Current =002013 First Expression - Time Domain.})
are the energy weights. This suggests employing a method akin to the
charge currents strategy. However, despite using the same assumption,
there are some subtle deviations from the reasoning for load currents,
as illustrated in Appendix \ref{sec: Appendix E - Energy current main formula.}.
The final formula is
\begin{align}
\dot{E}_{\alpha} & =-2\textrm{Re}\left\{ i\hbar\int\textrm{tr}\left[\mathcal{G}^{R}\left(t,t'\right)\partial_{t'}\varSigma_{\alpha}^{<}\left(t',t\right)\right.\right.\nonumber \\
 & \left.\left.+\mathcal{G}^{<}\left(t,t'\right)\partial_{t'}\varSigma_{\alpha}^{A}\left(t',t\right)\right]\right\} dt'.\label{eq: Energy Current =002013 Second Expression - Time Domain.}
\end{align}
Unlike Eq. (\ref{eq: Charge Current =002013 Second Expression - Time Domain.}),
the energy current formula shown above contains time derivatives over
self-energies. Up to this point, the expression for energy current
seems to be similar to the charge current. By applying the Wigner
transform and the gradient expansion to Eq. (\ref{eq: Energy Current =002013 Second Expression - Time Domain.})
we arrive at (see Appendix \ref{sec: Appendix F - Adiabatic expansion for energy current.})
\begin{align}
\dot{E}_{\alpha} & =-\frac{2}{h}\stackrel[N=0]{\infty}{\sum}\frac{\left(-1\right)^{N}}{N!}\int\varepsilon\textrm{Re}\left\{ \textrm{tr}\left\{ C_{\alpha,N}^{I}\right\} \right\} d\varepsilon\nonumber \\
 & +\frac{2}{h}\stackrel[N=0]{\infty}{\sum}\frac{\left(-1\right)^{N}}{N!}\int\textrm{Re}\left\{ \textrm{tr}\left\{ C_{\alpha,N}^{E}\right\} \right\} d\varepsilon.\label{eq: Energy Current =002013 Second Expression - Energy Domain.}
\end{align}

The additional term $C_{\alpha,N}^{E}$ is given by\footnote{Each of the $C_{\alpha,N}^{I}$ functions are dimensionless, whereas
the $C_{\alpha,N}^{E}$ functions have energy units.}
\begin{align*}
C_{\alpha,N}^{E} & =\left(\frac{i\hbar}{2}\right)^{N+1}\left(\partial_{T}^{N+1}\mathcal{\widetilde{G}}^{R}\partial_{\varepsilon}^{N}\varSigma_{\alpha}^{<}+\partial_{T}^{N+1}\mathcal{\widetilde{G}}^{<}\partial_{\varepsilon}^{N}\varSigma_{\alpha}^{A}\right).
\end{align*}

After introducing the second-order adiabatic expansion of the retarded
and lesser Green\textquoteright s functions into Eq. (\ref{eq: Energy Current =002013 Second Expression - Energy Domain.}),
we get the adiabatic expansion up to the second-order of the energy current
\begin{align}
\dot{E}_{\alpha} & \simeq\dot{E}_{\alpha}^{\left(0\right)}+\dot{E}_{\alpha}^{\left(1\right)}+\dot{E}_{\alpha}^{\left(1,1\right)}+\dot{E}_{\alpha}^{\left(2\right)}.\label{eq: Energy Current =002013 Second Order Adiabatic Expansion.}
\end{align}

The adiabatic energy current is 
\begin{align}
\dot{E}_{\alpha}^{\left(0\right)} & =\mathcal{\dot{\mathcal{E}}}_{\alpha}^{\left(0\right)}=\int\varepsilon\mathcal{K}_{\alpha}^{\left(0\right)}\left(T,\varepsilon\right)d\varepsilon,\label{eq: Energy Current =002013 Cero Order Adiabatic Term.}
\end{align}
and its first correction $\dot{E}_{\alpha}^{\left(1\right)}$ takes
the form
\begin{align}
\dot{E}_{\alpha}^{\left(1\right)} & =\underset{\nu}{\sum}\dot{\mathcal{E}}_{\alpha,\nu}^{\left(1\right)}\dot{X}_{\nu},\label{eq: Energy Current =002013 First Order Adiabatic Term.}
\end{align}
where
\begin{align*}
\mathcal{\dot{\mathcal{E}}}_{\alpha,\nu}^{\left(1\right)} & =\int\varepsilon\mathcal{K}_{\alpha,\nu}^{\left(1\right)}\left(T,\varepsilon\right)d\varepsilon.
\end{align*}

These terms are equivalent to those found previously by other authors,
see for example \onlinecite{ludovico2016Feb}. However, the last two terms
of Eq. (\ref{eq: Energy Current =002013 Second Order Adiabatic Expansion.})
represent original expressions, to our knowledge. They read as
\begin{align}
\dot{E}_{\alpha}^{\left(2\right)} & =\underset{\nu}{\sum}\mathcal{\dot{\mathcal{E}}}_{\alpha,\nu}^{\left(2\right)}\ddot{X}_{\nu},\label{eq: Energy Current =002013 Second Order Adiabatic Term =002013 Second Derivative.}\\
\dot{E}_{\alpha}^{\left(1,1\right)} & =\underset{\mu,\nu}{\sum}\mathcal{\dot{\mathcal{E}}}_{\alpha,\mu\nu}^{\left(1,1\right)}\dot{X}_{\mu}\dot{X}_{\nu},\label{eq: Energy Current =002013 Second Order Adiabatic Term =002013 First Cross Derivatives.}
\end{align}
where the energy fluxes take the form
\begin{align*}
\mathcal{\dot{\mathcal{E}}}_{\alpha,\nu}^{\left(2\right)} & =\int\varepsilon\mathcal{K}_{\alpha,\nu}^{\left(2\right)}\left(T,\varepsilon\right)d\varepsilon,\\
\mathcal{\dot{\mathcal{E}}}_{\alpha,\mu\nu}^{\left(1,1\right)} & =\int\varepsilon\mathcal{K}_{\alpha,\mu\nu}^{\left(1,1\right)}\left(T,\varepsilon\right)d\varepsilon.
\end{align*}

Note that, by comparing the adiabatic expansion of the charge current
Eq. (\ref{eq: Charge Current =002013 Second Order Adiabatic Expansion.})
with the energy one (Eq. \ref{eq: Energy Current =002013 Second Order Adiabatic Expansion.}),
we recover, order-by-order, the straightforward interpretation of
the energy current. It comes from the integration of all particles
entering and leaving a given lead, but multiplied by the energy each
particle carries. Mathematically this is a consequence of a subtle
cancellation of the second term of Eq. (\ref{eq: Energy Current =002013 Second Expression - Energy Domain.}).

\subsection{Heat current\label{subsec: Heat Current.}}

The heat current flowing through the $\alpha$-lead is the last observable
we will study up to the second order. We can determine the expectation
value of heat current based on the first law of thermodynamics \citep{esposito2015,bustosmarun2019}.
The total energy gained or lost by a lead $\alpha$ is the sum of
the heat and the work done by the particles being exchanged with the
local system. In this way, the heat current $J_{\alpha}$ is given
by
\begin{align}
J_{\alpha} & =\dot{E}_{\alpha}-\mu_{\alpha}\left\langle \dot{N}_{\alpha}\left(t\right)\right\rangle .\label{eq: Heat Current =002013 Definition.}
\end{align}

The formal definition of Eq. (\ref{eq: Heat Current =002013 Definition.})
contains two mean values that we developed in subsections \ref{subsec:Charge-current}
and \ref{subsec:Energy-current}. Using the previous results, we can
straightforwardly write the adiabatic expansion, up to the second order,
of the heat current, just as we have done before
\begin{align*}
J_{\alpha} & \simeq J_{\alpha}^{\left(0\right)}+J_{\alpha}^{\left(1\right)}+J_{\alpha}^{\left(1,1\right)}+J_{\alpha}^{\left(2\right)}.
\end{align*}

Employing Eqs. (\ref{eq: Charge Current =002013 Cero Order Adiabatic Term.}-\ref{eq: Charge Current =002013 Second Order Adiabatic Term =002013 First Cross Derivatives.})
and (\ref{eq: Energy Current =002013 Cero Order Adiabatic Term.}-\ref{eq: Energy Current =002013 Second Order Adiabatic Term =002013 First Cross Derivatives.}),
we found
\begin{align*}
J_{\alpha}^{\left(0\right)} & =\mathcal{J}_{\alpha}^{\left(0\right)}, & J_{\alpha}^{\left(2\right)} & =\underset{\nu}{\sum}\mathcal{J}_{\alpha,\nu}^{\left(2\right)}\ddot{X}_{\nu},\\
J_{\alpha}^{\left(1\right)} & =\underset{\nu}{\sum}\mathcal{J}_{\alpha,\nu}^{\left(1\right)}\dot{X}_{\nu}, & J_{\alpha}^{\left(1,1\right)} & =\underset{\mu,\nu}{\sum}\mathcal{J}_{\alpha,\mu\nu}^{\left(1,1\right)}\dot{X}_{\mu}\dot{X}_{\nu},
\end{align*}
where
\begin{align*}
\mathcal{J}_{\alpha}^{\left(0\right)} & =\mathcal{\dot{\mathcal{E}}}_{\alpha}^{\left(0\right)}-\mu_{\alpha}\mathcal{N}_{\alpha}^{\left(0\right)}, & \mathcal{\mathcal{J}}_{\alpha,\nu}^{\left(2\right)} & =\mathcal{\dot{\mathcal{E}}}_{\alpha,\nu}^{\left(2\right)}-\mu_{\alpha}\mathcal{N}_{\alpha,\nu}^{\left(2\right)},\\
\mathcal{\mathcal{J}}_{\alpha,\nu}^{\left(1\right)} & =\mathcal{\dot{\mathcal{E}}}_{\alpha,\nu}^{\left(1\right)}-\mu_{\alpha}\mathcal{N}_{\alpha,\nu}^{\left(1\right)}, & \mathcal{J}_{\alpha,\mu\nu}^{\left(1,1\right)} & =\mathcal{\dot{\mathcal{E}}}_{\alpha,\mu\nu}^{\left(1,1\right)}-\mu_{\alpha}\mathcal{N}_{\alpha,\mu\nu}^{\left(1,1\right)}.
\end{align*}

Likewise, the above results can be written as integrals, giving
\begin{align*}
\mathcal{\mathcal{J}}_{\alpha}^{\left(0\right)} & =\int\left(\varepsilon-\mu_{\alpha}\right)\mathcal{K}_{\alpha}^{\left(0\right)}\left(T,\varepsilon\right)d\varepsilon,\\
\mathcal{J}_{\alpha,\nu}^{\left(1\right)} & =\int\left(\varepsilon-\mu_{\alpha}\right)\mathcal{K}_{\alpha,\nu}^{\left(1\right)}\left(T,\varepsilon\right)d\varepsilon,\\
\mathcal{\mathcal{J}}_{\alpha,\nu}^{\left(2\right)} & =\int\left(\varepsilon-\mu_{\alpha}\right)\mathcal{K}_{\alpha,\nu}^{\left(2\right)}\left(T,\varepsilon\right)d\varepsilon,\\
\mathcal{J}_{\alpha,\mu\nu}^{\left(1,1\right)} & =\int\left(\varepsilon-\mu_{\alpha}\right)\mathcal{K}_{\alpha,\mu\nu}^{\left(1,1\right)}\left(T,\varepsilon\right)d\varepsilon.
\end{align*}

Thus far, we have offered a complete set of quantum transport observables
to understand how the second-order corrections work. These results
became a starting point to explore thermodynamic properties from a
quantum point of view.

\section{Models\label{sec: Models.}}

In the previous sections, we formulated the theory of adiabatic expansions
up to the second order for a set of fundamental observables. Now, we will
evaluate the charge, energy, and heat currents for two different time-dependent
devices. In studying these examples, we have three goals in mind.
First, we want to illustrate the usage of the developed formulas and
their utility for the thermodynamic analysis of nanomachines. Second,
we want to test their validity by showing that the results they produce
are physically reasonable. In particular, we will show that energy
and particle number are conserved at all times and for all parameters
in the order-by-order expansions \citep{bustosmarun2019}. This involves
the comparison of proven first-order formulas with the above-derived
second-order terms. The first two points are shown in the first analyzed
device, an atomic rotor in contact with two fixed leads (see Fig.
\ref{fig: Driven Atomic Rotor - Schematic Representation.}). The
second example consists of an oscillating quantum point contact (see
Fig. \ref{fig: Driven Quantum Point Contact - Schematic Representation.}).
There we want to illustrate the kind of phenomena that can arise from
the second-order terms of the observables. In this sense, we should
recall that traditional adiabatic quantum pumping \citep{brouwer1998},
requires the movement of at least two out-of-phase parameters. Instead,
the studied device shows monoparametric pumping thanks to the second-order
terms of charge currents, without requiring anharmonicities \citep{low2012}
or quantum interference effects mediated by magnetic fields \citep{foatorres2005}.

\subsection{Driven atomic rotor\label{subsec: Driven Atomic Rotor.}}

\begin{figure}[!h]
\begin{centering}
\includegraphics[width=2in]{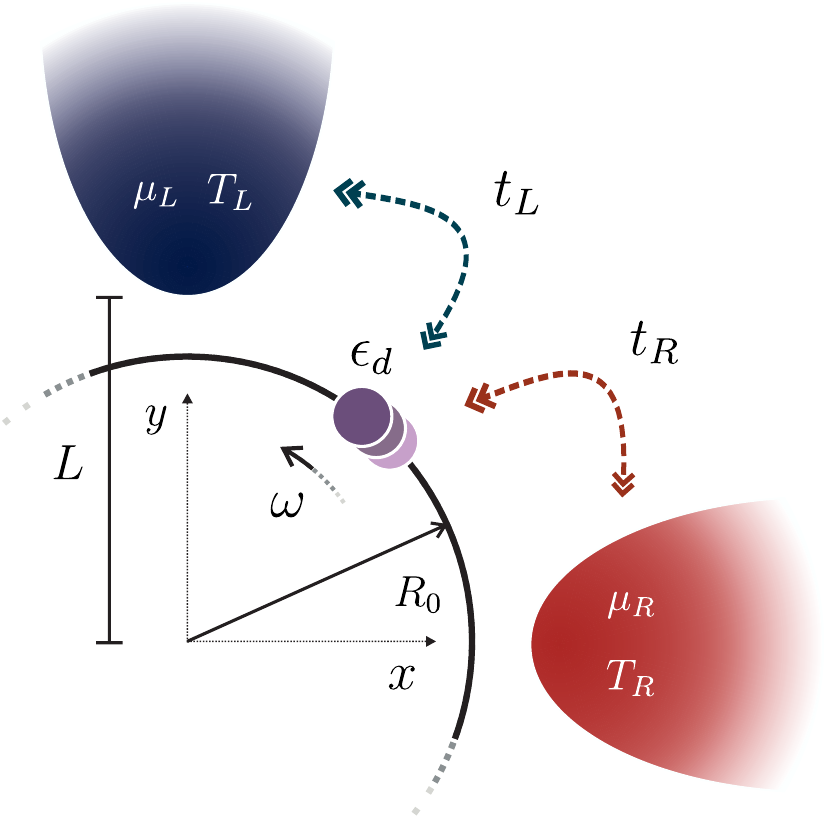}
\par\end{centering}
\caption{Representation of a bidimensional atomic rotor attached to two leads.
\label{fig: Driven Atomic Rotor - Schematic Representation.}}
\end{figure}

To model the atomic rotor we assume a minimal tight-binding model
consisting of a one-site system attached to two identical conduction
channels (the leads) through which an electrical current can flow.
We will name each of them as $L$ and $R$. The isolated site is linked
to the rotor and represents a quantum dot with a site energy $\epsilon_{d}$.
Two semi-infinite tight-binding chains set up the leads, where $t_{L}$
and $t_{R}$ are the tunneling between them and the rotor \citep{deghi2021},
as sketched in Fig. \ref{fig: Driven Atomic Rotor - Schematic Representation.}.

To keep matters simple, we assume only the hopping terms change with
time, having a linear dependence on the distance between the rotor
and the respective leads. Furthermore, we force the atomic rotor to
follow a circular trajectory with a fixed radius $R_{0}$ and a given
angular frequency $\omega$. Based on this outline, we arrive at (see
Appendix \ref{sec: Appendix G - Driven Atomic Rotor model.})
\begin{align}
t_{L} & =t_{\mathrm{max}}\left(1+a-a\sqrt{1+\beta\left(1-\sin\left(\omega t\right)\right)}\right),\label{eq: Driven Atomic Rotor - Tunnelling Between QD and L Lead - Linear Behaviour.}\\
t_{R} & =t_{\mathrm{max}}\left(1+a-a\sqrt{1+\beta\left(1-\cos\left(\omega t\right)\right)}\right),\label{eq: Driven Atomic Rotor - Tunnelling Between QD and R Lead - Linear Behaviour.}
\end{align}
where $t_{\mathrm{max}}$ is the maximum (in absolute value) tunneling
amplitude, $a$ sets the decay of the hopping terms with the distant
between the dot and the leads, and
\begin{align*}
\beta & =2\frac{R_{0}L}{\left(L-R_{0}\right)^{2}}.
\end{align*}

The description of the local system requires at least a three-site
Hamiltonian, which includes the site energy of the central dot and
the first sites of the leads. In this way, we circumvent the limitation
mentioned earlier of time-independent self-energies $\Sigma^{r,a,<}\left(\varepsilon\right)$.
The resulting adiabatic Green's functions $G^{r,a,<}$, whose explicit
forms can be found in \citep{deghi2021}, do not commute, in general,
with each other or $\Lambda$, as would be the case for a single site
local system; see the example of Ref. \onlinecite{kershaw2017}. Thus,
despite its simplicity, the analyzed example provides a challenging
test for the developed formulas.

Once we set up the model, we apply the charge current formulas. For
this purpose, we rewrite Eq. (\ref{eq: Charge Current =002013 Second Order Adiabatic Expansion.})
(see Appendix \ref{sec: Appendix G - Driven Atomic Rotor model.})
for the current over the $L$-lead as
\begin{align}
I_{L} & =\varOmega_{I_{L}}^{\left(0\right)}+\varOmega_{I_{L}}^{\left(1\right)}\omega+\left(\varOmega_{I_{L}}^{\left(1,1\right)}+\varOmega_{I_{L}}^{\left(2\right)}\right)\omega^{2}.\label{eq: Driven Atomic Rotor - Charge Current - Second Order Adiabatic Expansion.}
\end{align}
The term $\varOmega_{I_{L}}^{\left(0\right)}$ is the adiabatic contribution
given by Eq. (\ref{eq: Charge Current =002013 Cero Order Adiabatic Term.}),
$\varOmega_{I_{L}}^{\left(1\right)}$ is the first-order correction
dictated by Eq. (\ref{eq: Charge Current =002013 First Order Adiabatic Term.}),
and $\varOmega_{I_{L}}^{\left(1,1\right)}$ and $\varOmega_{I_{L}}^{\left(2\right)}$
stand for the second-order corrections of the charge current, given
by Eqs. (\ref{eq: Charge Current =002013 Second Order Adiabatic Term =002013 Second Derivative.})
and (\ref{eq: Charge Current =002013 Second Order Adiabatic Term =002013 First Cross Derivatives.})
respectively. 

The procedure used for the charge current can be extended to the remaining
studied observables. For example, the heat current formula takes the
form
\begin{align}
J_{L} & =\varOmega_{J_{L}}^{\left(0\right)}+\varOmega_{J_{L}}^{\left(1\right)}\omega+\left(\varOmega_{J_{L}}^{\left(1,1\right)}+\varOmega_{J_{L}}^{\left(2\right)}\right)\omega^{2}.\label{eq: Driven Atomic Rotor - Heat Current - Second Order Adiabatic Expansion.}
\end{align}

Despite their complexity, our goal is to ensure the validity of the
formulas. To achieve this point, we will use the order-by-order energy
conservation formulas for quantum transport \citep{bustosmarun2019}.
For our model, it reads
\begin{align}
Q_{J_{L}}^{\left(n\right)}-\frac{\delta\mu_{L}}{e}Q_{I_{L}}^{\left(n\right)}+Q_{J_{R}}^{\left(n\right)}-\frac{\delta\mu_{R}}{e}Q_{I_{R}}^{\left(n\right)}+W^{\left(n-1\right)} & =0.\label{eq: Driven Atomic Rotor - First Law of Thermodynamics - Order by Order.}
\end{align}
 The above formula holds for any integer $n$, where $n$ represents
the order of the adiabatic expansion of the quantity of interest,
$W^{\left(n-1\right)}$ is the $n-1$ order of the work done by the
rotor in a single cycle (see Appendix \ref{sec: Appendix I -Current Induced Forces.}),
$Q_{J_{\alpha}}^{\left(n\right)}$ is the heat pumped per revolution
of the rotor to the $\alpha$-lead, and $Q_{I_{\alpha}}^{\left(n\right)}$
is the pumped charge in a cycle. These quantities are evaluated as:
\begin{align}
W^{\left(n-1\right)} & =\stackrel[0]{\tau}{\int}\overrightarrow{F}^{\left(n-1\right)}\cdot\overrightarrow{X}dt\label{eq: Driven Atomic Rotor - Total Work per Cycle - Order by Order.}\\
Q_{J_{\alpha}}^{\left(n\right)} & =\stackrel[0]{\tau}{\int}J_{\alpha}^{\left(n\right)}dt,\label{eq: Driven Atomic Rotor - Total Charge Pumped per Cycle - Order by Order.}\\
Q_{I_{\alpha}}^{\left(n\right)} & =\stackrel[0]{\tau}{\int}I_{\alpha}^{\left(n\right)}dt.\label{eq: Driven Atomic Rotor - Total Heat Pumped per Cycle - Order by Order.}
\end{align}
Here, $J_{\alpha}^{\left(n\right)}$ and $I_{\alpha}^{\left(n\right)}$
are defined in Section \ref{sec: Observables.}, $\tau$ is the period
of the movement, and $\overrightarrow{F}^{\left(n-1\right)}$ is the
$n-1$ order of the adiabatic expansion of the electronic force $\overrightarrow{F}$.
The latter is a vector whose components are defined by
\begin{eqnarray*}
F_{\nu}\left(t\right) & = & -i\hbar\mathrm{Tr}\left[\Lambda_{\nu}\mathcal{G}^{<}\left(t,t\right)\right].
\end{eqnarray*}

It is important to highlight that the fulfillment of Eq. (\ref{eq: Driven Atomic Rotor - First Law of Thermodynamics - Order by Order.})
with second-order corrections of the heat and charge currents involves
the comparison of developed formulas with the well-known expressions
for the zero and first-order electronic forces, $\overrightarrow{F}^{(0)}$
and $\overrightarrow{F}^{(1)}$, which can be found in Refs. \citep{bode2011,deghi2021}.

A further magnitude of significance is the total energy pumped to
the $\alpha$-lead, which reads:
\begin{align*}
Q_{\dot{E}_{\alpha}}^{\left(n\right)} & =Q_{J_{\alpha}}^{\left(n\right)}-\frac{\delta\mu_{L}}{e}Q_{I_{\alpha}}^{\left(n\right)}.
\end{align*}
Rewriting Eq. (\ref{eq: Driven Atomic Rotor - First Law of Thermodynamics - Order by Order.})
using the above formula, we obtain a version of the first law of thermodynamics
which relates the total energy of electrons flowing between leads
and the mechanical work. However, in this version electron energy
and mechanical work are related through the different orders of their
adiabatic expansion.

Finally, charge conservation is another property that must be satisfied
for every order and can easily be tested in the present model
\begin{align*}
Q_{I_{L}}^{\left(n\right)}+Q_{I_{R}}^{\left(n\right)} & =0.
\end{align*}

\begin{figure}[!h]
\begin{centering}
\includegraphics[width=3in]{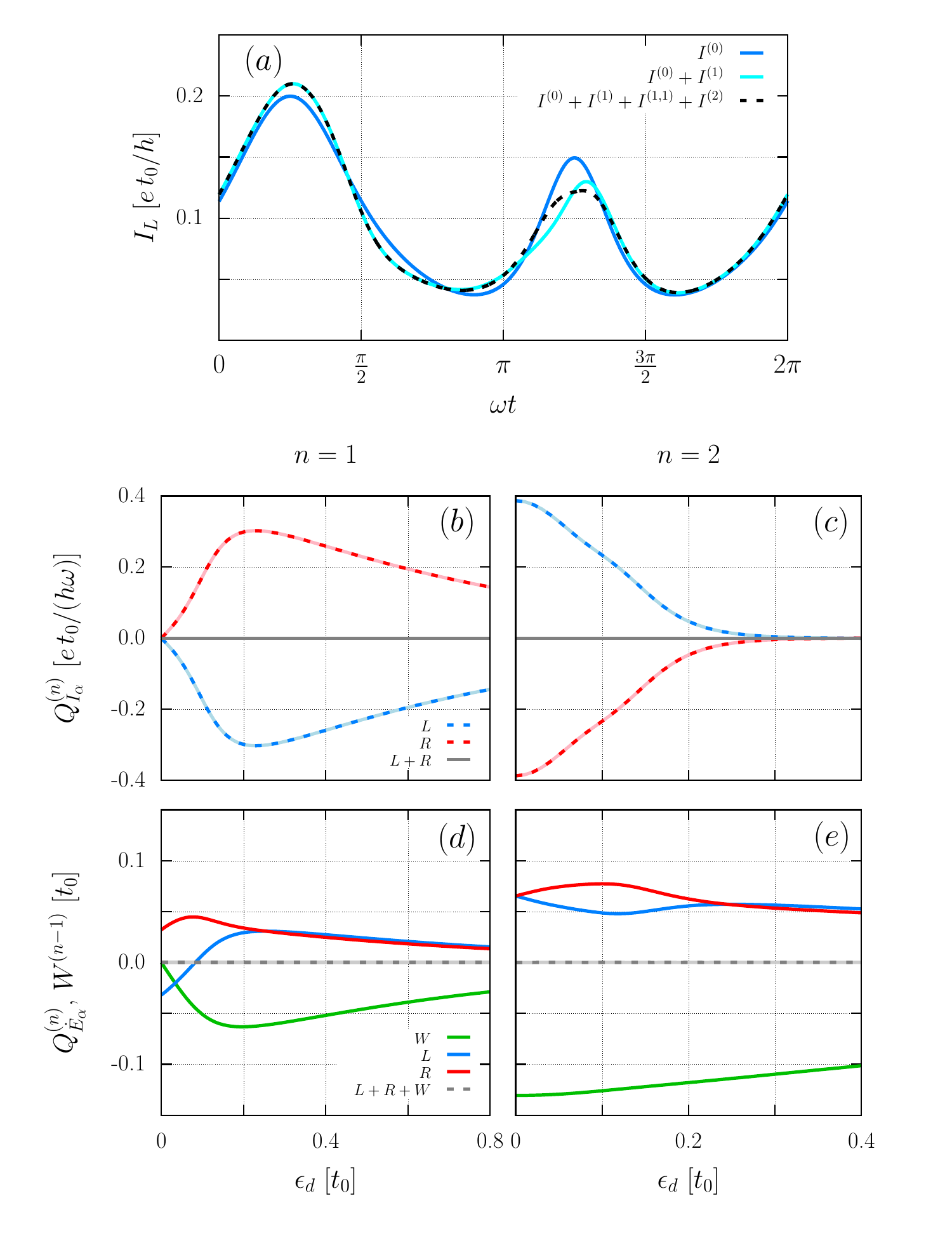}
\par\end{centering}
\caption{$\boldsymbol{(a)}$ \textbf{-} Charge current (in units of $e t_0/\hbar$) calculated up to different
orders as function of $\omega t$. The different lines are: $I^{(0)}$ for
blue, $I^{(0)}+I^{(1)}$ for cyan, and $I^{(0)}+I^{(1)}+I^{(2)}+I^{(1,1)}$
for the black dashed line. $\boldsymbol{(b)}$ and $\boldsymbol{(c)}$
\textbf{-} Pumped charge per cycle (in units of $e t_0/(\hbar\omega)$) for the first and second orders.
Red and blue dashed lines are $Q_{I_{R}}^{\left(n\right)}$ and $Q_{I_{L}}^{\left(n\right)}$
respectively, and the solid grey line is their sum. $\boldsymbol{(d)}$
and $\boldsymbol{(e)}$ \textbf{-} fulfillment of the order-by-order
energy conservation (in units of $t_0$). The blue and red lines are $Q_{\dot{E}_{\alpha}}^{\left(n\right)}$
for $\alpha=L$ and $\alpha=R$ respectively, the green line is $W^{(n-1)}$,
and the grey line shows the fulfillment of Eq. \ref{eq: Driven Atomic Rotor - First Law of Thermodynamics - Order by Order.}.
For all calculations, the temperatures are $k_{B}T_{L}=k_{B}T_{R}=0.01$,
the chemical potentials are $\delta\mu_{L}=0.1$ and $\delta\mu_{R}=-0.1$,
$R_{0}=0.5$, $a=0.44$, $L=1$ and $t_{\mathrm{max}}=1$. For Fig.
$(a)$, we fixed the quantum dot energy as $\epsilon_{d}=0.01$ and
$\omega=0.1$. \label{fig: Driven Atomic Rotor - Charge Current - Charge and Heat Pump per Cycle.}}
\end{figure}

Fig. (\ref{fig: Driven Atomic Rotor - Charge Current - Charge and Heat Pump per Cycle.})-($a$)
illustrates how the different orders of the charge current add up
over one period of the rotor. Figs. \ref{fig: Driven Atomic Rotor - Charge Current - Charge and Heat Pump per Cycle.}-($b$)
and \ref{fig: Driven Atomic Rotor - Charge Current - Charge and Heat Pump per Cycle.}-($c$)
verify that the studied model satisfies charge conservation at every
order for different values of the dot's energy $\epsilon_{d}$. Figures
\ref{fig: Driven Atomic Rotor - Charge Current - Charge and Heat Pump per Cycle.}-($d$)
and \ref{fig: Driven Atomic Rotor - Charge Current - Charge and Heat Pump per Cycle.}-($e$)
show, for different values of the dot's energy $\epsilon_{d}$, order-by-order
energy conservation for the first and second adiabatic corrections
of the heat and charge currents.

We can also use the present example to highlight the importance of
the developed formulas for the thermodynamic analysis of systems.
For example, in Fig. \ref{fig: Driven Atomic Rotor - Charge Current - Charge and Heat Pump per Cycle.}
we can see that there is charge pumping in both orders {[}see $(b)$
and $(c)${]} but energy pumping only occurs at first order. Note
that for $n=1$ {[}Fig. $(d)${]} and at $\epsilon_{d}$ close to
0, energy is on average going out of the left lead (blue line below
zero) and entering into the right lead (red line above zero). On the
contrary, at second-order of the currents {[}$n=2$, see $(e)${]},
energy coming from the external driving just dissipates through both
leads (red and blue curves are always positive). Here, $W^{(1)}$
{[}geen line of Fig. \ref{fig: Driven Atomic Rotor - Charge Current - Charge and Heat Pump per Cycle.}
$(e)${]} is the mechanical energy being dissipated by the electronic
friction, see section \ref{sec: Appendix I -Current Induced Forces.}
and Ref. \onlinecite{bustosmarun2019}. The negative value of $W^{(0)}$
in Fig. $(d)$ (the work per cycle done by CIFs) is due to the sign
choice of $\dot{X}\left(t\right)$, which in the present case implies
that the system is being forced to act as a pump, not as a motor,
see Ref. \onlinecite{bustosmarun2019}. As a final remark, we note that
for the present example second-order pumping points in a different
direction than the first-order one. This implies that even small deviations
from the adiabaticity should diminish the efficiency of adiabatic
quantum pumps.

\subsection{Oscillating QPC or STM's tip\label{subsec: Driven Quantum Point Contact.}}

\begin{figure}[!h]
\begin{centering}
\includegraphics[width=2in]{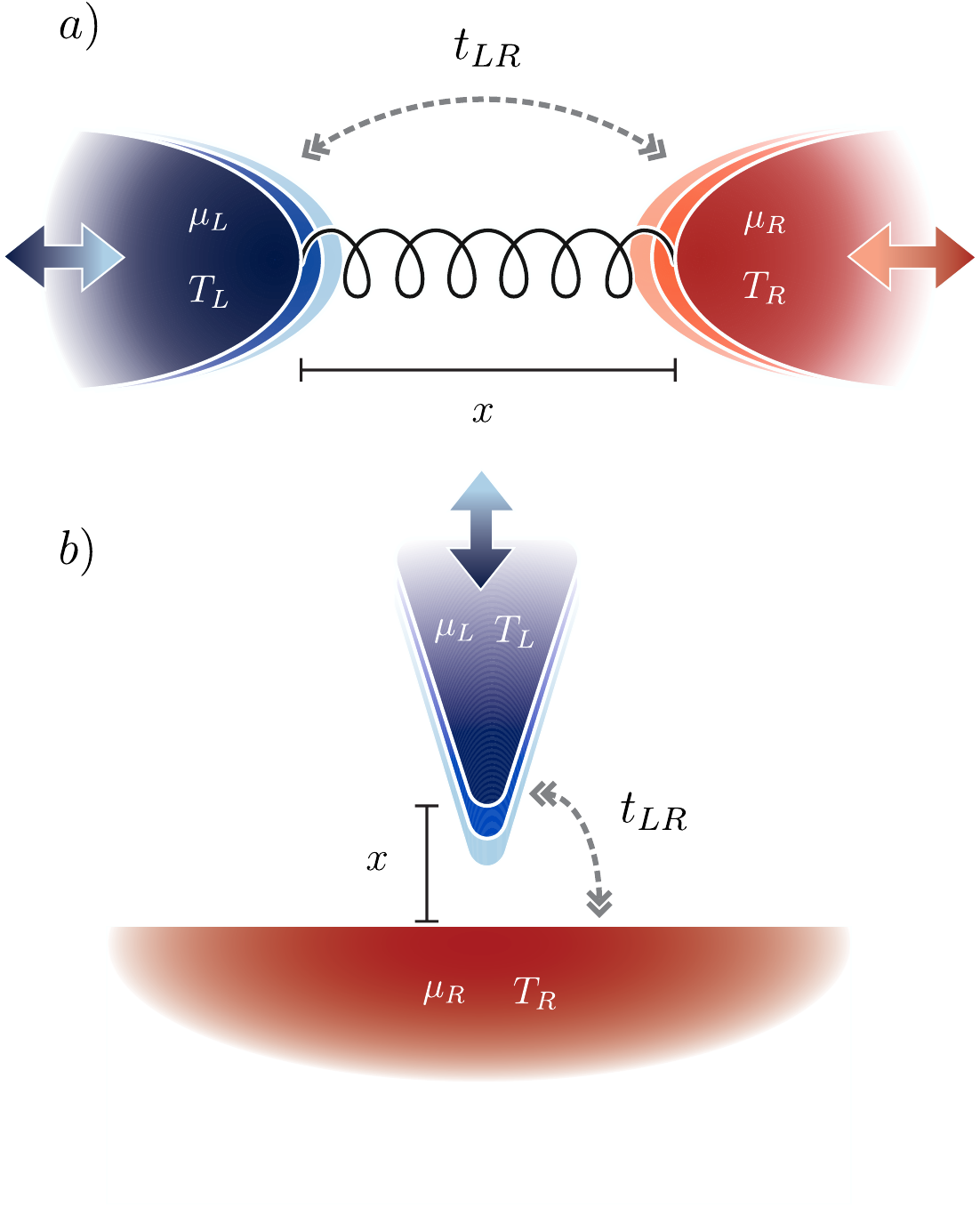}
\par\end{centering}
\caption{Scheme of two possible physical systems described by the model of
Sec. \ref{subsec: Driven Quantum Point Contact.}: $(a)$ an oscillating
quantum point contact and $(b)$ a oscillating tip of a scanning tunneling
microscope. \label{fig: Driven Quantum Point Contact - Schematic Representation.}}
\end{figure}

The system studied in this section consists of two conductors (not
necessarily of the same materials) close enough to let charge particles
tunnel from one to the other. Furthermore, the distance between the
conductors oscillates at a given frequency $\omega$. This may represent
different physical situations, which are in principle within experimental
possibilities \citep{chen2021}. Some of them are depicted in Fig.
\ref{fig: Driven Quantum Point Contact - Schematic Representation.}.

We modeled the different leads $L$ and $R$ using two distinct semi-infinite
homogeneous tight-binding chains, as in the previous example, each
characterized by a site energy and a hopping. The site energy of the
lead $L$ ($R$) is $\epsilon_{0L}$ ($\epsilon_{0R}$) and its hopping
is $t_{0L}$ ($t_{0R}$). The hopping between the leads $t_{LR}$
depends on their distance, exhibiting an exponential behavior (see
Appendix \ref{sec: Appendix H - Driven Quantum Point Contact Model.})
\begin{align}
t_{LR} & =t_{\mathrm{max}}e^{a\left(1-\frac{x}{x_{\mathrm{min}}}\right)},\label{eq: Driven Quantum Point Contact - Tunnelling Between L Lead and R Lead - Exponential Behaviour.}
\end{align}
where $t_{\mathrm{max}}$ is the maximum tunneling amplitude, $a$
is the decay factor, $x$ is the distance between the leads, and $x_{\mathrm{min}}$
is the minimum value of $x$. Assuming a forced oscillatory movement
and taking $\delta x+x_{\mathrm{min}}$ as the maximum length, the
distance between leads yields
\begin{align}
x & =x_{\mathrm{min}}+\frac{\delta x}{2}\left(1-\cos\left(\omega t\right)\right).\label{eq: Driven Quantum Point Contact - Distance Between L Lead and R Lead - Oscillating Behaviour.}
\end{align}

For the calculation of charge and heat currents, we rewrite the corresponding
formulas just as we did in Eqs. (\ref{eq: Driven Atomic Rotor - Charge Current - Second Order Adiabatic Expansion.})
and (\ref{eq: Driven Atomic Rotor - Heat Current - Second Order Adiabatic Expansion.}).
Moreover, for evaluating the fulfillment of the order-by-order energy
conservation in the present model, we will also resort to the same
definitions given in Eqs. (\ref{eq: Driven Atomic Rotor - First Law of Thermodynamics - Order by Order.}),
(\ref{eq: Driven Atomic Rotor - Total Work per Cycle - Order by Order.}),
and (\ref{eq: Driven Atomic Rotor - Total Heat Pumped per Cycle - Order by Order.}). 

\begin{figure}[!h]
\begin{centering}
\includegraphics[width=2.8in]{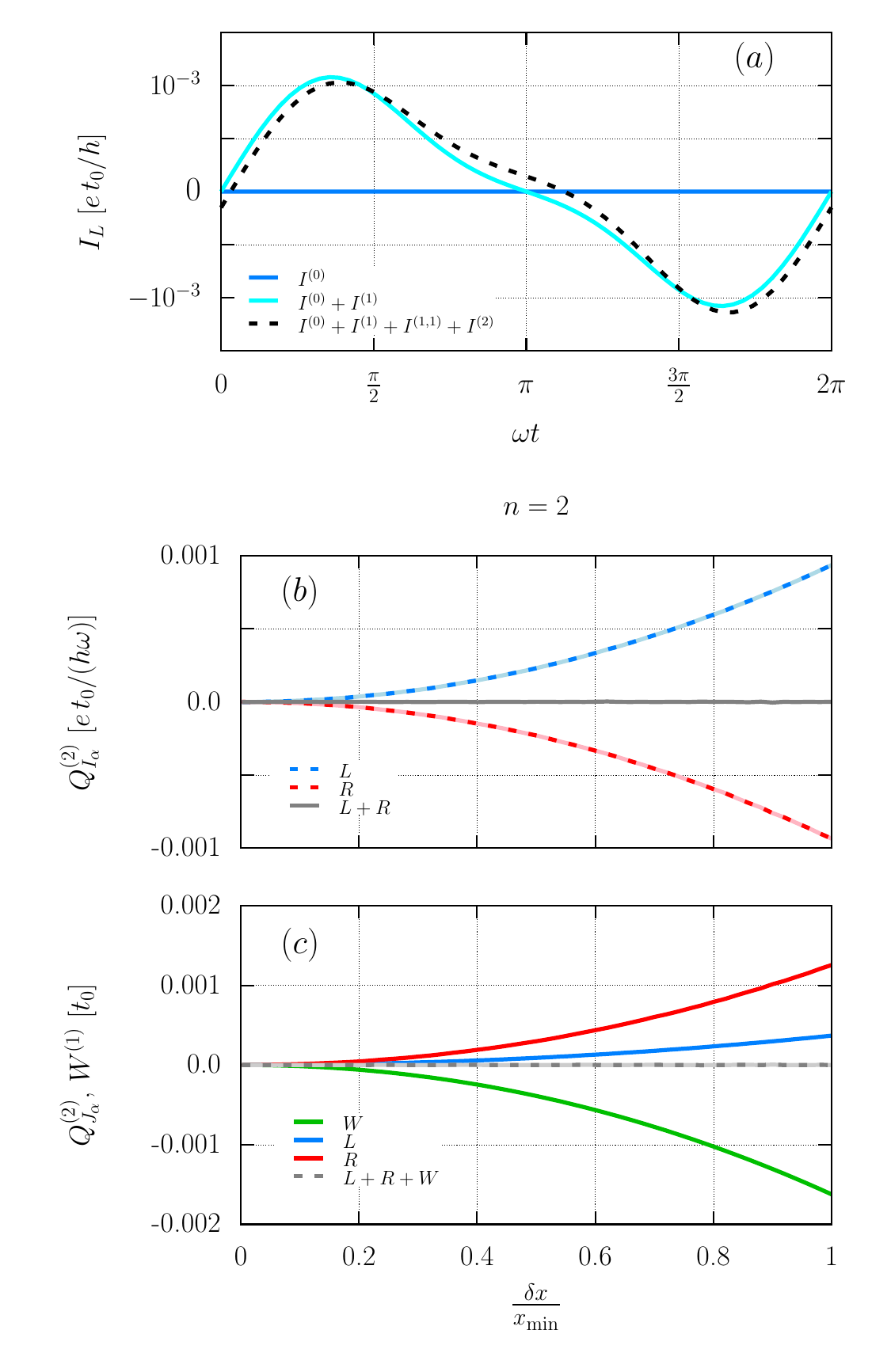}
\par\end{centering}
\caption{\textbf{$\boldsymbol{(a)}$ -} Charge current (in units of $e t_0/\hbar$) calculated up to different
orders as function of $\omega t$. The different lines are: $I^{(0)}$ for
blue, $I^{(0)}+I^{(1)}$ for cyan, and $I^{(0)}+I^{(1)}+I^{(2)}+I^{(1,1)}$
for the black dashed line. $\boldsymbol{(b)}$ \textbf{-} Second-order
($n=2$) pumped charges per cycle (in units of $e t_0/(\hbar \omega)$) as a function of $\delta x/x_{\mathrm{min}}$.
The blue dashed line belongs to the $L$-lead, the red one to the
$R$-lead, and the grey line is the sum of both contributions. $\boldsymbol{(c)}$
\textbf{-} fulfillment of the order-by-order energy conservation law (in units of $t_0$)
as a function of $\delta x/x_{\mathrm{min}}$. The blue and red lines
are $Q_{\dot{E}_{\alpha}}^{\left(n\right)}$ for $\alpha=L$ and $\alpha=R$
respectively, the green line is $W^{(n-1)}$, and the grey line shows
the fulfillment of Eq. \ref{eq: Driven Atomic Rotor - First Law of Thermodynamics - Order by Order.}.
In all the plots we used: $k_{B}T_{L}=k_{B}T_{R}=0.01$, $\delta\mu_{L}=\delta\mu_{R}=0$,
$t_{0L}=0.56$, $t_{0R}=1$, $\epsilon_{0L}=1$, $\epsilon_{0R}=0$,
$a=0.5$, and $t_{\mathrm{max}}=0.99$. In Fig. $(a)$ we fixed $\delta x/x_{\mathrm{min}}=0.5$
and $\omega=0.1$. \label{fig: Driven Quantum Point Contact - Charge Current - Charge and Heat Pump per Cycle.}}
\end{figure}

In Fig. \ref{fig: Driven Quantum Point Contact - Charge Current - Charge and Heat Pump per Cycle.}
we evaluate the current for a system in equilibrium, i.e. $\left(\mu_{L},T_{L}\right)=\left(\mu_{R},T_{R}\right)$,
with nonsymmetrical leads, which in this case means different energy
sites $\epsilon_{0L}\neq\epsilon_{0R}$. There, the zero-order heat
$J^{(0)}$ and charge currents $I^{(0)}$(adiabatic contributions)
are zero, as can be seen for $I^{(0)}$ in Fig. \ref{fig: Driven Quantum Point Contact - Charge Current - Charge and Heat Pump per Cycle.}-$(a)$
(blue solid line). Unlike the former example, the present one has
only one independent mechanical degree of freedom. This implies a
null contribution per cycle of the first-order heat and charge currents,
$J^{(1)}$ and $I^{(1)}$, see Ref. \onlinecite{brouwer1998}. Note in
Fig. \ref{fig: Driven Quantum Point Contact - Charge Current - Charge and Heat Pump per Cycle.}-$(a)$
that the cyan curve, $I^{(0)}+I^{(1)}$, is antisymmetric with respect
to $I=0$. However, the second-order term may allow particles and
heat to pump through the leads, even for a single-parameter system.
Note that, the dotted black line in Fig. \ref{fig: Driven Quantum Point Contact - Charge Current - Charge and Heat Pump per Cycle.}-$(a)$,
$I^{(0)}+I^{(1)}+I^{(2)}+I^{(1,1)}$, is not antisymmetric for $I=0$
but has a shift towards negative values. In this example, it is then
clear that second-order terms comprise the leading order for charge
and heat pumping, exhibiting an unexplored form of quantum pumping.

In Figs. \ref{fig: Driven Quantum Point Contact - Charge Current - Charge and Heat Pump per Cycle.}-$(b)$
and \ref{fig: Driven Quantum Point Contact - Charge Current - Charge and Heat Pump per Cycle.}-$(c)$,
we showed that the second-order term preserves, respectively, the
total charge and energy of the complete system (local system plus
the electron's leads). From this figure, it is also interesting to
note that, despite the charge current being pumped due to the second-order
terms {[}see (b){]}, there is no heat pumping. In Figs. \ref{fig: Driven Quantum Point Contact - Charge Current - Charge and Heat Pump per Cycle.}
(c) we can see that the heat per cycle $Q_{J}^{(2)}$ is positive
for left and right leads, i.e., the external work ($W$) is dissipated
as heat into the two leads. This mean that particles pumping from
the left lead,\footnote{According to Eq. (\ref{eq: Charge Current =002013 Definition.}),
the number of particles pumped per cycle is $-Q_{I_{L}}^{(2)}/e$.} does not compensate for the external work ($W$) being dissipated
as heat to this lead. This highlight, as in the previous example,
the usefulness of the developed formulas for thermodynamic analysis.

For a symmetrical system (identical leads), we found (not shown in
the manuscript) that no net particles move from one lead to another
in the former configuration, even for the second-order correction.
This behavior is expected since, due to inversion symmetry, there
is no reason for the current to go in a preferential direction. In
this sense, this also validates the second-order formulas.

\section{Conclusion\label{sec: Conclusion.}}

Throughout this work, we have presented expressions, based on NEGFs
within a Schwinger-Keldysh approach, for the second-order adiabatic
expansion of three different observables of quantum transport (energy,
heat, and charge currents). To our knowledge, NEGF second-order expansions
of heat and energy currents have not been treated before. Moreover,
the explicit formulas for the observables are general and ready to
be used in a wide range of systems. We only assumed that self-energies
are time-independent, which is naturally accomplished for most quantum
transport problems.

We have illustrated how the developed formulas produce physically
consistent results (including order-by-order energy and particle conservation
laws) using two simple models. We must emphasize that order-by-order
energy conservation strictly involves the comparison of the here-developed
expressions (second-order heat and charge currents) with a well-established
first-order expression (first-order current-induced force). Moreover,
we have shown in both examples how the developed formulas provide
useful thermodynamical information about the devices. Additionally,
in the second model, we have analyzed a phenomenon where second-order
monoparametric pumping takes place. This provides a glimpse of the
type of phenomenon that our expressions allow one to describe.

As for possible extensions of the present work, it would be interesting
a deeper exploration of second-order phenomena and their potential
use, as well as the development of similar formulas for other observables
like current-induced forces (also known as electron wind forces \citep{hoffmann2017})
or spin-transfer torque \citep{yang2021,xiao2021,wang2022,tang2024}.

\section{Acknowledgments}

This work was financially supported by Consejo Nacional de Investigaciones
Científicas y Técnicas (CONICET, PIP-2022-59241); Secretaría de Ciencia
y Tecnología de la Universidad Nacional de Córdoba (SECyT-UNC, Proyecto
Formar 2020); and Agencia Nacional de Promoción Científica y Tecnológica
(ANPCyT, PICT-2018-03587).

\appendix

\section{Wigner transform of Dyson\textquoteright s equation\label{sec: Appendix A - Wigner Transform of Dyson Equation.}}

Eq. (\ref{eq: Differential Dyson Equation for Green's Funtion.})
is the equation of motion of the retarded Green's function, usually
called Dyson's equation of motion. We can apply a Wigner transform
to this expression. The first term on the right-hand side of Eq. (\ref{eq: Differential Dyson Equation for Green's Funtion.})
is the Dirac delta, whose Wigner transform is the identity. Before
applying the Wigner transform to the second term, we need to rewrite
it as
\begin{align}
A\left(t,t'\right) & =\mathcal{G}^{R}\left(t,t'\right)H_{s}\left(t'\right)\nonumber \\
 & =\int\mathcal{G}^{R}\left(t,t_{2}\right)h\left(t_{2},t'\right)dt_{2},\label{eq: Appendix A - Motion Equation - Second Term.}
\end{align}
where
\begin{align*}
h\left(t_{2},t'\right) & =H_{S}\left(t_{2}\right)\delta\left(t_{2}-t'\right).
\end{align*}
Now, we can apply the Wigner transform to $A\left(t,t'\right)$, by
using the Moyal product \citep{maciejko2007}, yielding
\begin{align}
\widetilde{A} & =\mathcal{\widetilde{G}}^{R}\left(T,\varepsilon\right)e^{\frac{i\hbar}{2}\left(\overleftarrow{\partial}_{\varepsilon}\overrightarrow{\partial}_{T}-\overleftarrow{\partial}_{T}\overrightarrow{\partial}_{\varepsilon}\right)}\hat{H}_{S}\left(T\right).\label{eq: Appendix A - Moyal Product - Second Term.}
\end{align}
where we used $\widetilde{h}\left(T,\varepsilon\right)=H_{s}\left(T\right)$.
The exponential operator is interpreted as usual in terms of the Maclaurin
series of the exponential function, i.e., $e^{\hat{o}}=\sum_{n}\frac{\hat{o}^{n}}{n!}$
The arrows in the exponential indicate the direction in which the
derivatives should be applied. For example, the symbol $\overrightarrow{\partial}_{\varepsilon}$
means: derivative with respect to $\varepsilon$ of the function to
the right, $\hat{H}_{S}$ in the above case, while $\overleftarrow{\partial}_{\varepsilon}$
is the same but the derivative should be applied to the function to
the left, $\mathcal{\widetilde{G}}^{R}\left(T,\varepsilon\right)$
in the above case.

The third term of Eq. (\ref{eq: Differential Dyson Equation for Green's Funtion.}),
\begin{align*}
B\left(t,t'\right) & =\int\mathcal{G}^{R}\left(t,t_{1}\right)\Sigma^{R}\left(t_{1},t'\right)dt_{1},
\end{align*}
can also be Wigner transformed by using the Moyal product, giving
\begin{align}
\widetilde{B} & =\mathcal{\widetilde{G}}^{R}\left(T,\varepsilon\right)e^{\frac{i\hbar}{2}\left(\overleftarrow{\partial}_{\varepsilon}\overrightarrow{\partial}_{T}-\overleftarrow{\partial}_{T}\overrightarrow{\partial}_{\varepsilon}\right)}\Sigma^{R}\left(\varepsilon\right).\label{eq: Appendix A - Moyal Product - Third Term.}
\end{align}

To deal with the right-hand side of Eq. (\ref{eq: Differential Dyson Equation for Green's Funtion.})
we need to first apply the chain rule,

\begin{align*}
\partial_{t'}\mathcal{G}^{R} & =\frac{1}{2}\partial_{T}\mathcal{G}^{R}-\partial_{\tau}\mathcal{G}^{R},
\end{align*}
where $T$ and $\tau$ are defined in Eqs. (\ref{eq: Wigner Transform - Slow Coordinate.})
and (\ref{eq: Wigner Transform - Fast Coordinate.}) respectively.
Then, after Wigner transform it we obtained
\begin{align*}
\widetilde{\left(\partial_{t'}\mathcal{G}^{r}\right)} & =\frac{1}{2}\partial_{T}\mathcal{\widetilde{G}}^{R}\left(T,\varepsilon\right)+\frac{i\varepsilon}{\hbar}\mathcal{\widetilde{G}}^{R}\left(T,\varepsilon\right),
\end{align*}
where we used the fact that Green\textquoteright s functions have
compact support in the energy domain. The final result of all the
above is Eq. (\ref{eq: Retarded Green Funtion - Moyal Product.})
of the main text.

In Eqs. (\ref{eq: Appendix A - Moyal Product - Second Term.}) and
(\ref{eq: Appendix A - Moyal Product - Third Term.}) we used a compact
notation, based on a exponential operator. However, the same equations
can also be written as two infinite sums
\begin{align*}
\widetilde{A} & =\stackrel[N=0]{\infty}{\sum}\frac{1}{N!}\left(\frac{i\hbar}{2}\right)^{N}C_{N}^{A}\left(T,\varepsilon\right),
\end{align*}
and
\begin{align*}
\widetilde{B} & =\stackrel[N=0]{\infty}{\sum}\frac{1}{N!}\left(\frac{i\hbar}{2}\right)^{N}C_{N}^{B}\left(T,\varepsilon\right),
\end{align*}
where
\begin{align}
C_{N}^{A} & =\stackrel[j=0]{N}{\sum}\left(-1\right)^{N-j}\binom{N}{j}\partial_{T}^{N-j}\partial_{\varepsilon}^{j}\mathcal{\widetilde{G}}^{R}\partial_{T}^{j}\partial_{\varepsilon}^{N-j}H_{s},\label{eq: Appendix A - Moyal Product - Auxiliar Second Term.}\\
C_{N}^{B} & =\stackrel[j=0]{N}{\sum}\left(-1\right)^{N-j}\binom{N}{j}\partial_{T}^{N-j}\partial_{\varepsilon}^{j}\mathcal{\widetilde{G}}^{R}\partial_{T}^{j}\partial_{\varepsilon}^{N-j}\Sigma^{R}.\label{eq: Appendix A - Moyal Product - Auxiliar Third Term.}
\end{align}
On the one hand, the slow time $T$ only affects the local system,
according to our assumption. Therefore, the dynamics of the leads
only depends on the energy $\varepsilon$. On the other hand, the
Hamiltonian of the local system $H_{s}$ only depends on $T$ and
not on $\varepsilon$. Then
\begin{align}
\partial_{\varepsilon}^{N-j}H_{s} & =0\;\forall N-1>j\geq0.\label{eq: Appendix A - Local System Property - Energy Derivative.}\\
\partial_{T}^{j}\Sigma^{R} & =0\;\forall N\geq j>1.\label{eq: Appendix A - Reservoir Property - Time Derivative.}
\end{align}
This can be used in Eq. (\ref{eq: Retarded Green Funtion - Moyal Product.})
of the main text to obtain a simpler expression. The results is
\begin{align}
1 & =-\frac{i\hbar}{2}\partial_{T}\mathcal{\widetilde{G}}^{R}+\varepsilon\mathcal{\widetilde{G}}^{R}-\stackrel[N=0]{\infty}{\sum}\frac{1}{N!}\left(\frac{i\hbar}{2}\right)^{N}C_{N}.\label{eq: Appendix A - Retarded Green's Function - Moyal Product.}
\end{align}
where we have introduced Eqs. (\ref{eq: Appendix A - Local System Property - Energy Derivative.})
and (\ref{eq: Appendix A - Reservoir Property - Time Derivative.})
into Eqs. (\ref{eq: Appendix A - Moyal Product - Auxiliar Second Term.})
and (\ref{eq: Appendix A - Moyal Product - Auxiliar Third Term.})
respectively, and we have defined
\begin{align*}
C_{N} & =\left(-1\right)^{N}\partial_{T}^{N}\mathcal{\widetilde{G}}^{R}\partial_{\varepsilon}^{N}\Sigma^{R}+\partial_{\varepsilon}^{N}\mathcal{\widetilde{G}}^{R}\partial_{T}^{N}H_{s}.
\end{align*}
The zero and first order terms of last expression can be compared
with Eq. 20 of Ref. \onlinecite{bode2012Jan}.

\section{Adiabatic expansion for retarded Green's function\label{sec: Apendix B - Adiabatic Expansion for Retarded Green's Functions.}}

The adiabatic retarded Green function, given in Eq. (\ref{eq: Adiabatic Retarded Green Function}),
satisfies the condition
\begin{align}
\left[G^{R}\left(T,\varepsilon\right)\right]^{-1} & =\varepsilon I-H_{s}\left(T\right)-\Sigma^{R}\left(T,\varepsilon\right).\label{eq: Appendix B - Adiabatic Retarded Green's Function - Inverse.}
\end{align}
Then, the derivative of $G^{R}$ with respect to $\varepsilon$ gives
\begin{align}
\partial_{\varepsilon}G^{R} & =-G^{r}\left(1-\partial_{\varepsilon}\Sigma^{R}\right)G^{r}.\label{eq: Appendix B - Adiabatic Retarded Green's Function - Derivative.}
\end{align}
where we used $\partial A=-A\partial\left[A^{-1}\right]A$.

Next, we have to change the sequence of the terms in the Eq. (\ref{eq: Appendix A - Retarded Green's Function - Moyal Product.})
to introduce the retarded Green's function and its derivative, given
in Eqs. (\ref{eq: Adiabatic Retarded Green Function}) and (\ref{eq: Appendix B - Adiabatic Retarded Green's Function - Derivative.}).
Then, using the slow-time derivative, Eqs. (\ref{eq: First Slow Time Derivative - Local Hamiltonian.}),
and multiplying the adiabatic retarded Green\textquoteright s function
on both sides of the obtained equation, the resulting expression takes
the form
\begin{align}
\mathcal{\widetilde{G}}^{R} & =G^{R}+\frac{i\hbar}{2}\stackrel[\nu=1]{M}{\sum}\partial_{\varepsilon}\Sigma^{R}\varLambda_{\nu}G^{R}\dot{X}_{\nu}\label{eq: Appendix B - Retarded Green's Function - Adiabatic Expansion.}\\
 & -\frac{i\hbar}{2}\partial_{T}\mathcal{\widetilde{G}}^{R}\left[G^{R}\right]^{-1}\partial_{\varepsilon}G^{R}+\stackrel[N=2]{\infty}{\sum}\frac{1}{N!}\left(\frac{i\hbar}{2}\right)^{N}C_{N}.\nonumber 
\end{align}
Importantly, no approximations have been made in Eq. (\ref{eq: Appendix B - Retarded Green's Function - Adiabatic Expansion.}).
Therefore, as long as the adiabatic expansion is valid (the series
is convergent), the equation provides the exact evolution of $\mathcal{\widetilde{G}}^{R}$
along a path driven by the mechanical degrees of freedom $X_{\mu}$.
However, note that $\mathcal{\widetilde{G}}^{R}$ (the exact retarded
Green's function) depends in turns on the time derivative of $\mathcal{\widetilde{G}}^{R}$,
which is not known. In principle, the only retarded Green's function
known in advance is the adiabatic one, $G^{r}$. To solve this, one
applies the iterative method twice. This consists of replacing $\mathcal{\widetilde{G}}^{R}$
on the right hand side by a first approximation ($G^{r}$), then using
the resulting Green function to improve the approximation of $\mathcal{\widetilde{G}}^{R}$
on the right hand side, and so on. During this iterative process one
identify terms of different orders given by the order of the derivatives
with respect to $X_{\mu}$ and cut the infinite series at a given
order (second order in our case). Note that the order also coincide
with the exponent of $\hbar$ or the exponent of the driving frequency
$\omega$ for the case $X_{\mu}\propto\cos\left(\omega t\right)$.
By using this method we arrive, after some algebra, at Eq. (\ref{eq: Adiabatic Expansion of Retarded Green's Function.})
of the main text.

\section{Adiabatic expansion for lesser Green's function\label{sec: Appendix C - Adiabatic Expansion for Lesser Green's Functions.}}

Starting from Eq. (\ref{eq: Lesser Green Funtion - Moyal Product.}),
which gives the lesser Green\textquoteright s function Wigner transforms
in close notation, we can write its gradient expansion as a double
infinite sum
\begin{align}
\mathcal{\widetilde{G}}^{<} & =\stackrel[N,K=0]{\infty}{\sum}\frac{1}{N!K!}\left(\frac{i\hbar}{2}\right)^{N+K}W_{N,K}^{<}\left(T,\varepsilon\right),\label{eq: Appendix C - Lesser Green's Function - Moyal Product Serie Expansion - Extended.}
\end{align}
where we have introduced the nested definition
\begin{align*}
W_{N,K}^{<} & =\stackrel[j=0]{N}{\sum}\stackrel[l=0]{K}{\sum}\left(-1\right)^{N}\binom{N}{j}\binom{K}{l}\partial_{T}^{N-j}\partial_{\varepsilon}^{j}\mathcal{\widetilde{G}}^{R}C_{j,l}^{N,K}
\end{align*}
and
\begin{align*}
C_{j,l}^{N,K} & =\left(-1\right)^{j+l}\partial_{T}^{j}\partial_{\varepsilon}^{N-j}\left\{ \partial_{T}^{K-l}\partial_{\varepsilon}^{l}\Sigma^{<}\partial_{T}^{l}\partial_{\varepsilon}^{K-l}\mathcal{\widetilde{G}}^{A}\right\} .
\end{align*}
The aforementioned expression is obtained by straightforwardly making
use of the Moyal product twice.

The lesser self-energy is not time-dependent, since we have assumed
that the retarded self-energy is not controlled by the slow time.
This implies
\begin{align}
\partial_{T}^{K-l}\Sigma^{<} & =0\;\forall N-l\neq0,\label{eq: Appendix C - Reservoir Property - Time Derivative - 1.}\\
\partial_{T}^{j-m}\Sigma^{<} & =0\;\forall j-m\neq0.\label{eq: Appendix C - Reservoir Property - Time Derivative - 2.}
\end{align}
Putting Eqs. (\ref{eq: Appendix C - Reservoir Property - Time Derivative - 1.})
and (\ref{eq: Appendix C - Reservoir Property - Time Derivative - 2.})
in Eq. (\ref{eq: Appendix C - Lesser Green's Function - Moyal Product Serie Expansion - Extended.}),
result in the simplified expression
\begin{align}
\mathcal{\widetilde{G}}^{<} & =\stackrel[Q=0]{\infty}{\sum}\frac{1}{Q!}\left(\frac{i\hbar}{2}\right)^{Q}W_{Q}^{<}\left(T,\varepsilon\right),\label{eq: Appendix C - Lesser Green's Function - Moyal Product Serie Expansion - Shorted.}
\end{align}
where
\begin{align*}
W_{Q}^{<} & =\stackrel[q=0]{Q}{\sum}\stackrel[s=0]{q}{\sum}\stackrel[j=0]{Q-q}{\sum}\left(-1\right)^{q}\binom{Q}{q+j}\binom{q+j}{j}\binom{q}{s}C_{q,j}^{Q},
\end{align*}
and
\begin{align*}
C_{q,j}^{Q} & =\partial_{T}^{q}\partial_{\varepsilon}^{j}\mathcal{\widetilde{G}}^{R}\partial_{\varepsilon}^{Q-\left(j+s\right)}\Sigma^{<}\partial_{\varepsilon}^{s}\partial_{T}^{Q-q}\mathcal{\widetilde{G}}^{A}.
\end{align*}
The zero and first-order terms of Eq. (\ref{eq: Appendix C - Lesser Green's Function - Moyal Product Serie Expansion - Shorted.})
coincide with Eq. 25 of Ref. \onlinecite{bode2012Jan} and Eq. 3 of Ref.
\onlinecite{bode2011}, once the iterative method explained in the previous
section is applied. The result up to the second order is shown in
Eq. (\ref{eq: Adiabatic Expansion of Lesser Green's Function.}) of
the main text.

\section{Adiabatic expansion for charge currents\label{sec: Appendix D - Adiabatic expansion for particle's current.}}

We start by rewriting Eq. (\ref{eq: Charge Current =002013 Second Expression - Time Domain.})
as

\begin{align}
I_{\alpha} & =2e\textrm{Re}\left\{ \textrm{tr}\left[F_{\alpha}\left(t\right)\right]\right\} ,\label{eq: Appendix D - Charge Current - Main Formula - Time Domain.}
\end{align}
where the auxiliary operator $F_{\alpha}$ is defined by
\begin{align}
F_{\alpha} & =\int\left[\mathcal{G}^{R}\left(t,t'\right)\varSigma_{\alpha}^{<}\left(t',t\right)\right.\label{eq: Appendix D - Charge Current - F Term - Time Domain.}\\
 & \left.+\mathcal{G}^{<}\left(t,t'\right)\varSigma_{\alpha}^{A}\left(t',t\right)\right]dt'\nonumber 
\end{align}
Now, our goal is to transform the $F_{\alpha}$ operator to the energy
domain and then apply the inverse Wigner transform. This allows us
to recover the charge current in the time domain written as an integral
with a kernel in the energy domain. To carry out his technique, we
start with the relationship between $F_{\alpha}$ and its Wigner transform
$\widetilde{F}_{\alpha}$, whose formula takes the form
\begin{align}
F_{\alpha} & =\frac{1}{2\pi\hbar}\int\widetilde{F}_{\alpha}\left(T,\varepsilon\right)d\varepsilon.\label{eq: Appendix D - Wigner Inverse Transform - F Term - Time Domain.}
\end{align}

The next step is to put the gradient expansion in effect over Eq.
(\ref{eq: Appendix D - Charge Current - F Term - Time Domain.}),
leading to
\begin{align}
\widetilde{F}_{\alpha} & =\stackrel[N=0]{\infty}{\sum}\stackrel[j=0]{N}{\sum}\frac{1}{N!}\left(-\frac{i\hbar}{2}\right)^{N}\binom{N}{j}C_{\alpha,N,j}^{F}\left(T,\varepsilon\right),\label{eq: Appendix D - Wigner Transform - F Term - First Formulation - Energy Domain.}
\end{align}
where
\begin{align*}
C_{\alpha,N,j}^{F} & =\left(-1\right)^{j}\partial_{T}^{N-j}\partial_{\varepsilon}^{j}\mathcal{\widetilde{G}}^{R}\partial_{T}^{j}\partial_{\varepsilon}^{N-j}\varSigma_{\alpha}^{<}\\
 & +\left(-1\right)^{j}\partial_{T}^{N-j}\partial_{\varepsilon}^{j}\mathcal{\widetilde{G}}^{<}\partial_{T}^{j}\partial_{\varepsilon}^{N-j}\varSigma_{\alpha}^{A}.
\end{align*}

As we have mentioned, we assumed that the leads are unaffected by
the slow time variation. For such reason. Therefore, the following
rules hold
\begin{align}
\partial_{T}^{j}\varSigma_{\alpha}^{A} & =0\;\forall N\geq j\geq1,\label{eq: Appendix D - Reservoir Property - Time Derivative - 1.}\\
\partial_{T}^{j}\varSigma_{\alpha}^{<} & =0\;\forall N\geq j\geq1.\label{eq: Appendix D - Reservoir Property - Time Derivative - 2.}
\end{align}
Putting Eqs. (\ref{eq: Appendix D - Reservoir Property - Time Derivative - 1.})
and (\ref{eq: Appendix D - Reservoir Property - Time Derivative - 2.})
in Eq. (\ref{eq: Appendix D - Wigner Transform - F Term - First Formulation - Energy Domain.}),
the Wigner transform of the $F_{\alpha}$ operator reads
\begin{align}
\widetilde{F}_{\alpha} & =\stackrel[N=0]{\infty}{\sum}\frac{1}{N!}\left(-\frac{i\hbar}{2}\right)^{N}C_{\alpha,N}^{F}\left(T,\varepsilon\right),\label{eq: Appendix D - Wigner Transform - F Term - Second Formulation - Energy Domain.}
\end{align}
where
\begin{align*}
C_{\alpha,N}^{F} & =\partial_{T}^{N}\mathcal{\widetilde{G}}^{R}\partial_{\varepsilon}^{N}\varSigma_{\alpha}^{<}+\partial_{T}^{N}\mathcal{\widetilde{G}}^{<}\partial_{\varepsilon}^{N}\varSigma_{\alpha}^{A}.
\end{align*}
To get the desired result, we must first plug the last equation into
Eq. (\ref{eq: Appendix D - Wigner Inverse Transform - F Term - Time Domain.})
and then into Eq. (\ref{eq: Appendix D - Charge Current - Main Formula - Time Domain.}),
leading to the formula outlined in Eq. (\ref{eq: Charge Current =002013 Second Expression - Energy Domain.}).

\section{Energy current main formula\label{sec: Appendix E - Energy current main formula.}}

Taking Eq. (\ref{eq: Energy Current =002013 First Expression - Time Domain.})
as a starting point, we intend to get an expression of the energy
current in terms of self-energies, which lets us identify the leads.
The quoted formula can be read as the product of the hooping between
the local system and the leads, with those elements of the lesser
Green\textquoteright s function that connect the local system and
leads, given by Eq. (\ref{eq: Lesser Green's Funtion - Formal Definition - System betwen Lead.}).
However, we want to express the energy current in terms of the Green's
functions and self-energies of the local system only, Eq. (\ref{eq: Lesser Green's Funtion - Formal Definition - System betwen System.}).
For this purpose, we will use the Keldysh technique and the Langreth
theorem of analytic continuation to achieve this goal \citep{maciejko2007,vanLeeuwen2006}.
The mean result is
\begin{align*}
J_{\alpha} & =-2\textrm{Re}\left\{ \int\mathcal{J}_{\alpha}\left(t,t'\right)dt'\right\} ,
\end{align*}
where
\begin{align*}
\mathcal{J}_{\alpha} & =\underset{l,s}{\sum}\left(\mathcal{G}_{l,s}^{R}\left(t,t'\right)f_{\alpha,s,l}^{<}\left(t',t\right)+\mathcal{G}_{l,s}^{<}\left(t,t'\right)f_{\alpha,s,l}^{A}\left(t',t\right)\right).
\end{align*}
Here, $\mathcal{G}_{l,s}^{A}$ are the elements of the retarded Green\textquoteright s
function given in Eq. (\ref{eq: Retarded Green's Funtion - Formal Definition.}),
and $\mathcal{G}_{l,s}^{<}$ are the lesser Green\textquoteright s
function specified in Eq. (\ref{eq: Lesser Green's Funtion - Formal Definition - System betwen System.}).
In addition, we have introduced the following auxiliary functions
\begin{align}
f_{\alpha,s,l}^{<} & =\underset{k}{\sum}\epsilon_{\alpha k}t_{s,\alpha k}g_{\alpha k}^{<}\left(t',t\right)t_{\alpha k,l},\label{eq: Appendix E - Self Energy Auxiliar Term - Lesser.}\\
f_{\alpha,s,l}^{A} & =\underset{k}{\sum}\epsilon_{\alpha k}t_{s,\alpha k}g_{\alpha k}^{A}\left(t',t\right)t_{\alpha k,l}.\label{eq: Appendix E - Self Energy Auxiliar Term - Advenced.}
\end{align}

As you can notice, these propagators satisfy the differential equations:
\begin{align}
\left(i\hbar\frac{\partial}{\partial t'}-\epsilon_{\alpha k}\right)g_{\alpha k}^{<}\left(t',t\right) & =0,\label{eq: Appendix E - One Particle Propagator - Lesser Term.}\\
\left(i\hbar\frac{\partial}{\partial t'}-\epsilon_{\alpha k}\right)g_{\alpha k}^{A}\left(t',t\right) & =\delta\left(t'-t\right).\label{eq: Appendix E - One Particle Propagator - Advanced Term.}
\end{align}
Moreover, and following our assumptions, we also have:
\begin{align*}
\partial_{t'}\left(t_{s,\alpha k}t_{\alpha k,l}\right) & =0.
\end{align*}
Then, putting Eqs. (\ref{eq: Appendix E - One Particle Propagator - Lesser Term.})
and (\ref{eq: Appendix E - One Particle Propagator - Advanced Term.})
into Eqs. (\ref{eq: Appendix E - Self Energy Auxiliar Term - Lesser.})
and (\ref{eq: Appendix E - Self Energy Auxiliar Term - Advenced.}),
respectively, and using the previous condition, we arrive at
\begin{align*}
f_{\alpha,s,l}^{<} & =\partial_{t'}\varSigma_{\alpha,s,l}^{<}\left(t',t\right),\\
f_{\alpha,s,l}^{A} & =\partial_{t'}\varSigma_{\alpha,s,l}^{A}\left(t',t\right)+\underset{k}{\sum}t_{s,\alpha k}\delta\left(t'-t\right)t_{\alpha k,l}.
\end{align*}
Note that we have used the definitions of the self-energies given
in Eqs. (\ref{eq: Alpha Lead Advanced Self Energy - Time Domain.})
and (\ref{eq: Alpha Lead Lesser Self Energy - Time Domain.}). The
last equations allow us to write the following
\begin{align*}
\mathcal{J}_{\alpha} & =\mathcal{J}_{\alpha}^{\left(0\right)}\left(t,t'\right)+\mathcal{J}_{\alpha}^{\left(1\right)}\left(t,t'\right),
\end{align*}
where
\begin{align*}
\mathcal{J}_{\alpha}^{\left(A\right)} & =\textrm{tr}\left\{ \mathcal{G}^{R}\left(t,t'\right)\varSigma_{\alpha}^{0}\right\} \delta\left(t'-t\right),\\
\mathcal{J}_{\alpha}^{\left(B\right)} & =\textrm{tr}\left\{ \mathcal{G}^{R}\left(t,t'\right)\partial_{t'}\varSigma_{\alpha}^{<}\left(t',t\right)+\mathcal{G}^{<}\left(t,t'\right)\partial_{t'}\varSigma_{\alpha}^{A}\left(t',t\right)\right\} .
\end{align*}
For compactness, we have defined above the operator
\begin{align*}
\varSigma_{\alpha,s,l}^{0} & =\underset{k}{\sum}t_{s,\alpha k}t_{\alpha k,l}.
\end{align*}

With this, the energy current takes the form
\begin{align*}
J_{\alpha} & =J_{\alpha}^{\left(A\right)}+J_{\alpha}^{\left(B\right)},
\end{align*}
where
\begin{align*}
J_{\alpha}^{\left(A\right)} & =-2\textrm{Re}\left\{ \int\mathcal{J}_{\alpha}^{\left(A\right)}\left(t,t'\right)dt'\right\} ,\\
J_{\alpha}^{\left(B\right)} & =-2\textrm{Re}\left\{ \int\mathcal{J}_{\alpha}^{\left(B\right)}\left(t,t'\right)dt'\right\} .
\end{align*}
However, with the aid of the properties $\left(\mathcal{G}^{<}\right)^{\text{\ensuremath{\dagger}}}=-\mathcal{G}^{<}$
and $\left(\varSigma_{\alpha}^{0}\right)^{\dagger}=\varSigma_{\alpha}^{0}$,
the following can be proved
\begin{align*}
J_{\alpha}^{\left(A\right)} & =0.
\end{align*}
Therefore, the total energy current is set only by $J_{\alpha}^{\left(A\right)}$,
resulting in Eq. (\ref{eq: Energy Current =002013 Second Expression - Time Domain.}).

\section{Adiabatic expansion for energy currents\label{sec: Appendix F - Adiabatic expansion for energy current.}}

In this section, we apply a method analogous to the one used for the
charge current. We start by defining the auxiliary operator
\begin{align}
M_{\alpha} & =\int\left[\mathcal{G}^{R}\left(t,t'\right)\partial_{t'}\varSigma_{\alpha}^{<}\left(t',t\right)\right.\nonumber \\
 & \left.+\mathcal{G}^{<}\left(t,t'\right)\partial_{t'}\varSigma_{\alpha}^{A}\left(t',t\right)\right]dt'\label{eq: Appendix F - Energy Current - M Term - Time Domain.}
\end{align}
This enables us to write the energy current as
\begin{align}
J_{\alpha} & =-2\textrm{Re}\left\{ i\hbar\textrm{tr}\left[M_{\alpha}\left(t\right)\right]\right\} .\label{eq: Appendix F - Energy Current - Main Formula - Time Domain.}
\end{align}

Once again, the procedure is to apply the Wigner transform to the
$M_{\alpha}$ operator and afterward the inverse transform. Then,
the gradient expansion of Eq. (\ref{eq: Appendix F - Energy Current - M Term - Time Domain.})
takes the form
\begin{align}
\widetilde{M}_{\alpha} & =\stackrel[N=0]{\infty}{\sum}\stackrel[j=0]{N}{\sum}\frac{1}{N!}\left(-\frac{i\hbar}{2}\right)^{N}\binom{N}{j}C_{\alpha,N,j}^{M}\left(T,\varepsilon\right),\label{eq: Appendix F - Wigner Transform - M Term - First Formulation - Energy Domain.}
\end{align}
where
\begin{align*}
C_{\alpha,N,j}^{M} & =\left(-1\right)^{j}\partial_{T}^{N-j}\partial_{\varepsilon}^{j}\mathcal{\widetilde{G}}^{R}\partial_{T}^{j}\partial_{\varepsilon}^{N-j}\widetilde{\partial_{t'}\varSigma_{\alpha}^{<}}\\
 & +\left(-1\right)^{j}\partial_{T}^{N-j}\partial_{\varepsilon}^{j}\mathcal{\widetilde{G}}^{<}\partial_{T}^{j}\partial_{\varepsilon}^{N-j}\widetilde{\partial_{t'}\varSigma_{\alpha}^{A}}.
\end{align*}
where $\widetilde{\partial_{t'}\varSigma_{\alpha}}$ is the Wigner
transform of $\partial_{t'}\varSigma_{\alpha}\left(t',t\right)$.

The charge current formula, shown in Eq. (\ref{eq: Appendix D - Wigner Transform - F Term - First Formulation - Energy Domain.}),
can be compared with the above transform. The main difference between
both lies in the self-energies, where the one given in Eq. (\ref{eq: Appendix F - Wigner Transform - M Term - First Formulation - Energy Domain.})
implies the Wigner transform of the self-energy time derivative. To
find the expressions for these terms, we just need to use the definitions
of the Wigner coordinates, given in Eqs. (\ref{eq: Wigner Transform - Slow Coordinate.})
and (\ref{eq: Wigner Transform - Fast Coordinate.}), followed by
the chain rule. We then apply the Wigner transform to the resulting
expressions and assume that the self-energies, which are solely energy-dependent
($\varSigma_{\alpha}\left(\varepsilon\right)\equiv\varSigma_{\alpha}$),
vanish at high and low energies. The result is
\begin{align}
\widetilde{\partial_{t'}\varSigma_{\alpha}^{<}} & =\frac{1}{2}\partial_{T}\varSigma_{\alpha}^{<}-i\frac{\varepsilon}{\hbar}\varSigma_{\alpha}^{<},\label{eq: Appendix F - Lesser Self Energy Time Derivative - Wigner Transform - Chain Rule.}\\
\widetilde{\partial_{t'}\varSigma_{\alpha}^{A}} & =\frac{1}{2}\partial_{T}\varSigma_{\alpha}^{A}-i\frac{\varepsilon}{\hbar}\varSigma_{\alpha}^{A}.\label{eq: Appendix F - Advanced Self Energy Time Derivative  - Wigner Transform - Chain Rule.}
\end{align}

Then, plugging the Eqs. (\ref{eq: Appendix F - Lesser Self Energy Time Derivative - Wigner Transform - Chain Rule.})
and (\ref{eq: Appendix F - Advanced Self Energy Time Derivative  - Wigner Transform - Chain Rule.})
into Eq. (\ref{eq: Appendix F - Wigner Transform - M Term - First Formulation - Energy Domain.}),
together with the conditions of Eqs. (\ref{eq: Appendix D - Reservoir Property - Time Derivative - 1.})
and (\ref{eq: Appendix D - Reservoir Property - Time Derivative - 2.}),
the Wigner transform of $M_{\alpha}$ can be written as
\begin{align}
\widetilde{M}_{\alpha} & =-\frac{1}{2}\stackrel[N=0]{\infty}{\sum}\frac{1}{N!}\left(-\frac{i\hbar}{2}\right)^{N-1}C_{\alpha,N}^{M}\left(T,\varepsilon\right),\label{eq: Appendix F - Wigner Transform - M Term - Second Formulation - Energy Domain.}
\end{align}
where
\begin{align*}
C_{\alpha,N}^{M} & =\partial_{T}^{N}\mathcal{\widetilde{G}}^{R}\partial_{\varepsilon}^{N}\left(\varepsilon\varSigma_{\alpha}^{<}\right)+\partial_{T}^{N}\mathcal{\widetilde{G}}^{<}\partial_{\varepsilon}^{N}\left(\varepsilon\varSigma_{\alpha}^{A}\right).
\end{align*}
We can apply the chain rule to $\partial_{\varepsilon}^{N}\left(\varepsilon\varSigma_{\alpha}^{<}\right)$
giving
\begin{align}
\partial_{\varepsilon}^{N}\left(\varepsilon\varSigma_{\alpha}^{<}\right) & =\varepsilon\partial_{\varepsilon}^{N}\varSigma_{\alpha}^{<}+N\partial_{\varepsilon}^{N-1}\varSigma_{\alpha}^{<},\label{eq: Appendix F - Lesser Self Energy Energy Derivative - Chain Rule.}\\
\partial_{\varepsilon}^{N}\left(\varepsilon\varSigma_{\alpha}^{A}\right) & =\varepsilon\partial_{\varepsilon}^{N}\varSigma_{\alpha}^{A}+N\partial_{\varepsilon}^{N-1}\varSigma_{\alpha}^{A}.\label{eq: Appendix F - Advanced Self Energy Energy Derivative - Chain Rule.}
\end{align}
These results allow us to rewrite the $\widetilde{M}_{\alpha}$ of
Eq. (\ref{eq: Appendix F - Wigner Transform - M Term - Second Formulation - Energy Domain.})
in a closed form
\begin{align}
\widetilde{M}_{\alpha} & =-\frac{i\varepsilon}{\hbar}\widetilde{F}_{\alpha}-\frac{1}{2}\stackrel[N=0]{\infty}{\sum}\frac{1}{N!}\left(-\frac{i\hbar}{2}\right)^{N}C_{\alpha,N}\left(T,\varepsilon\right),\label{eq: Appendix F - Wigner Transform - M Term - Third Formulation - Energy Domain.}
\end{align}
where the term $\widetilde{F}_{\alpha}$ is that defined for the charge
current (see Eq. (\ref{eq: Appendix D - Wigner Transform - F Term - Second Formulation - Energy Domain.})),
and
\begin{align*}
C_{\alpha,N} & =\partial_{T}^{N+1}\mathcal{\widetilde{G}}^{R}\partial_{\varepsilon}^{N}\varSigma_{\alpha}^{<}+\partial_{T}^{N+1}\mathcal{\widetilde{G}}^{<}\partial_{\varepsilon}^{N}\varSigma_{\alpha}^{A}.
\end{align*}

Finally, we need to apply the inverse Wigner transform to Eq. (\ref{eq: Appendix F - Wigner Transform - M Term - Third Formulation - Energy Domain.})
and then put this into the energy current definition {[}Eq. (\ref{eq: Appendix F - Energy Current - Main Formula - Time Domain.}){]}.
The result is Eq. (\ref{eq: Energy Current =002013 Second Expression - Energy Domain.}).

\section{Driven atomic rotor model\label{sec: Appendix G - Driven Atomic Rotor model.}}

In subsection \ref{subsec: Driven Atomic Rotor.}, we provide a brief
account of the fundamental constituents of the atomic rotor model.
We will use two semi-infinite tight-binding chains to represent the
leads attached to the quantum dot and include the first site of each
chain as part of the local system. Then, the Hamiltonian of the local
system reads:
\begin{align*}
\mathbf{H}_{S} & =\begin{pmatrix}\epsilon_{0} & -t_{L} & 0\\
-t_{L} & \epsilon_{d} & -t_{R}\\
0 & -t_{R} & \epsilon_{0}
\end{pmatrix}.
\end{align*}

The adiabatic retarded Green's function, given by Eq. (\ref{eq: Adiabatic Retarded Green Function}),
takes the form:
\begin{align}
\mathbf{G}^{r} & =\left(\varepsilon\mathbf{I}-\mathbf{H}_{S}-\boldsymbol{\mathbf{\Sigma}}_{L}^{R}-\boldsymbol{\mathbf{\Sigma}}_{R}^{R}\right)^{-1},\label{eq: Appendix G - Retarded Adiabatic Green Function - Quantum Dot Attache Two Leads.}
\end{align}
where the self-energies coming from the decimation of the left and
right leads ($\boldsymbol{\mathbf{\Sigma}}_{L}^{R}$ and $\mathbf{\boldsymbol{\Sigma}}_{R}^{R}$
respectively) are given by
\begin{align*}
\boldsymbol{\mathbf{\Sigma}}_{L}^{R} & =\begin{pmatrix}\Sigma_{0}^{R}\left(\varepsilon\right) & 0 & 0\\
0 & 0 & 0\\
0 & 0 & 0
\end{pmatrix}, & \boldsymbol{\mathbf{\Sigma}}_{R}^{R} & =\begin{pmatrix}0 & 0 & 0\\
0 & 0 & 0\\
0 & 0 & \Sigma_{0}^{R}\left(\varepsilon\right)
\end{pmatrix}.
\end{align*}
Here, $\Sigma_{0}^{R}$ is the self-energy (in the energy domain)
of a semi-infite tight-binding chain. It is often expressed as
\begin{align}
\Sigma_{0}^{R}\left(\varepsilon\right) & =\Delta_{0}\left(\varepsilon\right)-i\Gamma_{0}\left(\varepsilon\right),\label{eq: Appendix G - Self Energy - Homogeneous Semi-infinity Tight-Binding Chain.}
\end{align}
where
\begin{align*}
\Delta_{0} & =t_{0}\begin{cases}
\frac{\varepsilon-\epsilon_{0}}{2t_{0}}-\sqrt{\left(\frac{\varepsilon-\epsilon_{0}}{2t_{0}}\right)^{2}-1} & \frac{\varepsilon-\epsilon_{0}}{2t_{0}}\geq1,\\
\frac{\varepsilon-\epsilon_{0}}{2t_{0}} & 1\geq\frac{\varepsilon-\epsilon_{0}}{2t_{0}}\geq-1,\\
\frac{\varepsilon-\epsilon_{0}}{2t_{0}}+\sqrt{\left(\frac{\varepsilon-\epsilon_{0}}{2t_{0}}\right)^{2}-1} & -1\geq\frac{\varepsilon-\epsilon_{0}}{2t_{0}},
\end{cases}\\
\Gamma_{0} & =t_{0}\begin{cases}
0 & \frac{\varepsilon-\epsilon_{0}}{2V_{0}}\geq1,\\
\sqrt{1-\left(\frac{\varepsilon-\epsilon_{0}}{2t_{0}}\right)^{2}} & 1\geq\frac{\varepsilon-\epsilon_{0}}{2t_{0}}\geq-1,\\
0 & -1\geq\frac{\varepsilon-\epsilon_{0}}{2t_{0}}.
\end{cases}
\end{align*}
Above, $\epsilon_{0}$ and $t_{0}$ are, respectively, the site energy
and the hopping between neighboring sites of the tight-binding chain.

In our model, the mechanical DOF affects only $t_{L}$ and $t_{R}$,
which can be taken as generalized coordinates related to the position
of the rotor $\theta$ by some function. Then, the matrix operator
$\boldsymbol{\Lambda}_{\theta}$ can be written as

\begin{eqnarray*}
\boldsymbol{\varLambda}_{\theta} & = & \frac{\partial t_{L}}{\partial\theta}\boldsymbol{\varLambda}_{t_{L}}+\frac{\partial t_{R}}{\partial\theta}\boldsymbol{\varLambda}_{t_{R}}.
\end{eqnarray*}
Here, the matrix operators $\boldsymbol{\mathbf{\varLambda}}_{t_{L}}$
and $\boldsymbol{\mathbf{\varLambda}}_{t_{R}}$, necessary to calculate
the charge and heat currents (see sections \ref{subsec:Charge-current}
and \ref{subsec: Heat Current.}), as well as the electronic force
(see Section \ref{sec: Appendix I -Current Induced Forces.}), are:

\begin{align*}
\boldsymbol{\varLambda}_{t_{L}} & =-\begin{pmatrix}0 & 1 & 0\\
1 & 0 & 0\\
0 & 0 & 0
\end{pmatrix}\quad\mathrm{and} & \boldsymbol{\varLambda}_{t_{R}} & =-\begin{pmatrix}0 & 0 & 0\\
0 & 0 & 1\\
0 & 1 & 0
\end{pmatrix}.
\end{align*}

So far, we have provided a general tight-binding model for a quantum
dot connected with two leads. The next stage requires the specification
of the geometric parameters of the atomic rotor. Typically the dependence
of hopping parameters with the distance are modeled by exponential
functions. In our case, that would mean taking
\begin{align}
t_{L} & =t_{m}e^{a\left(1-\frac{\Delta r_{L}}{r_{0L}}\right)},\label{eq: Appendix G - Tunneling Between QD and L Lead - Exponential Behaviour.}\\
t_{R} & =t_{m}e^{a\left(1-\frac{\Delta r_{R}}{r_{0R}}\right)}.\label{eq: Appendix G - Tunneling Between QD and R Lead - Exponential Behaviour.}
\end{align}
where $\Delta r_{L}$ and $\Delta r_{R}$ are the spatial distances
between the quantum dot and the closest chain sites to the $L$ and
$R$ leads, respectively, while $r_{0L}$ and $r_{0R}$ are the smallest
values of $\Delta r_{L}$ and $\Delta r_{R}$ along the trajectory.
Using polar coordinates, we have
\begin{align}
\Delta r_{L} & =\sqrt{L^{2}+R_{0}^{2}-2R_{0}L\sin\left(\theta\right)},\label{eq: Appendix G - Spatial Distance Between QD to Site L =002013 Polar Coordinates.}\\
\Delta r_{R} & =\sqrt{L^{2}+R_{0}^{2}-2R_{0}L\cos\left(\theta\right)}.\label{eq: Appendix G - Spatial Distance Between QD to Site R =002013 Polar Coordinates.}
\end{align}
where $\theta$, $L$, and $R_{0}$ are defined in Fig. \ref{fig: Driven Atomic Rotor - Schematic Representation.},
and $r_{0L}$ and $r_{0R}$ satisfy the condition
\begin{align}
r_{0L} & =r_{0R}=L-R_{0}.\label{eq: Appendix G - Closest Spatial Distance Between QD to the Sites L and R =002013 Cartesian Coordinates.}
\end{align}

Plugging the Eqs. (\ref{eq: Appendix G - Spatial Distance Between QD to Site L =002013 Polar Coordinates.}),
(\ref{eq: Appendix G - Spatial Distance Between QD to Site R =002013 Polar Coordinates.})
and (\ref{eq: Appendix G - Closest Spatial Distance Between QD to the Sites L and R =002013 Cartesian Coordinates.})
into Eqs. (\ref{eq: Appendix G - Tunneling Between QD and L Lead - Exponential Behaviour.})
and (\ref{eq: Appendix G - Tunneling Between QD and R Lead - Exponential Behaviour.})
make the model available to evaluate charge and heat currents. However,
since our purpose is to outline typical behaviors of the currents
and not to focus on the particularities of the used models, we decided
to use a linearized version of Eqs. (\ref{eq: Appendix G - Tunneling Between QD and L Lead - Exponential Behaviour.})
and (\ref{eq: Appendix G - Tunneling Between QD and R Lead - Exponential Behaviour.}).
In this way, we arrive at Eqs. (\ref{eq: Driven Atomic Rotor - Tunnelling Between QD and L Lead - Linear Behaviour.})
and (\ref{eq: Driven Atomic Rotor - Tunnelling Between QD and R Lead - Linear Behaviour.}).

Taking into account the model, we are now able to put the charging
current calculation into action. For a fixed radius and angular position
given by $\theta=\omega t$, we rewrite Eqs. (\ref{eq: Charge Current =002013 Cero Order Adiabatic Term.}),
(\ref{eq: Charge Current =002013 First Order Adiabatic Term.}), (\ref{eq: Charge Current =002013 Second Order Adiabatic Term =002013 Second Derivative.})
and (\ref{eq: Charge Current =002013 Second Order Adiabatic Term =002013 First Cross Derivatives.})
as
\begin{align*}
I_{L}^{\left(0\right)} & =\varOmega_{I_{L}}^{\left(0\right)}, & I_{L}^{\left(1,1\right)} & =\varOmega_{I_{L}}^{\left(1,1\right)}\omega^{2},\\
I_{L}^{\left(1\right)} & =\varOmega_{I_{L}}^{\left(1\right)}\omega, & I_{L}^{\left(2\right)} & =\varOmega_{I_{L}}^{\left(2\right)}\omega^{2},
\end{align*}
where
\begin{align*}
\varOmega_{I_{L}}^{\left(1\right)} & =-e\left(\mathcal{N}_{L,t_{L}}^{\left(1\right)}\frac{\partial t_{L}}{\partial\theta}+\mathcal{N}_{L,t_{R}}^{\left(1\right)}\frac{\partial t_{R}}{\partial\theta}\right),\\
\varOmega_{I_{L}}^{\left(2\right)} & =-e\left(\mathcal{N}_{L,t_{L}}^{\left(2\right)}\frac{\partial^{2}t_{L}}{\partial\theta^{2}}+\mathcal{N}_{L,t_{R}}^{\left(2\right)}\frac{\partial^{2}t_{R}^{2}}{\partial\theta^{2}}\right),\\
\varOmega_{I_{L}}^{\left(1,1\right)} & =-e\left(\mathcal{N}_{L,t_{L}t_{L}}^{\left(1,1\right)}\left(\frac{\partial t_{L}}{\partial\theta}\right)^{2}+\mathcal{N}_{L,t_{R}t_{R}}^{\left(1,1\right)}\left(\frac{\partial t_{R}}{\partial\theta}\right)^{2}\right)\\
 & -e\left(\mathcal{N}_{L,t_{L}t_{R}}^{\left(1,1\right)}+\mathcal{N}_{L,t_{R}t_{L}}^{\left(1,1\right)}\right)\frac{\partial t_{L}}{\partial\theta}\frac{\partial t_{R}}{\partial\theta}.
\end{align*}

Similar expressions can be obtained for the heat currents, see Sections
\ref{subsec:Energy-current} and \ref{subsec: Heat Current.}.

\section{Driven quantum point contact model\label{sec: Appendix H - Driven Quantum Point Contact Model.}}

In subsection \ref{subsec: Driven Quantum Point Contact.}, we discussed
the physical behavior of this kind of device and provided a brief
introduction to its modeling via two tight-binding semi-infinite chains
with tunneling between them. From this starting point, our next step
is to give the Hamiltonian of the local system, which reads
\begin{align*}
\mathbf{H}_{S} & =\begin{pmatrix}\epsilon_{0L} & -t_{LR}\\
-t_{LR} & \epsilon_{0R}
\end{pmatrix}.
\end{align*}

The adiabatic retarded Green's function assumes the same form as Eq.
(\ref{eq: Appendix G - Retarded Adiabatic Green Function - Quantum Dot Attache Two Leads.}),
The self-energies in this case are
\begin{align*}
\boldsymbol{\mathbf{\Sigma}}_{L}^{R} & =\begin{pmatrix}\Sigma_{0L}^{R}\left(\varepsilon\right) & 0\\
0 & 0
\end{pmatrix}, & \boldsymbol{\mathbf{\Sigma}}_{R}^{R} & =\begin{pmatrix}0 & 0\\
0 & \Sigma_{0R}^{R}\left(\varepsilon\right)
\end{pmatrix}.
\end{align*}
where $\Sigma_{0L}^{r}$ ($\Sigma_{0R}^{r}$) is given by Eq. (\ref{eq: Appendix G - Self Energy - Homogeneous Semi-infinity Tight-Binding Chain.})
but replacing the site energy $\epsilon_{0}$ with $\epsilon_{0L}$
($\epsilon_{0R}$) and the hopping $t_{0}$ by $t_{0L}$ ($t_{0R}$).

Since in this system only one parameter moves with time, we have a
single $\boldsymbol{\mathbf{\varLambda}}_{\nu}$ matrix given by
\begin{align*}
\boldsymbol{\mathbf{\varLambda}} & =-\begin{pmatrix}0 & 1\\
1 & 0
\end{pmatrix}.
\end{align*}
Finally, the dependence of the hopping with the separation between
leads is given in Eqs. (\ref{eq: Driven Quantum Point Contact - Tunnelling Between L Lead and R Lead - Exponential Behaviour.})
and (\ref{eq: Driven Quantum Point Contact - Distance Between L Lead and R Lead - Oscillating Behaviour.}).

\section{Electronic forces\label{sec: Appendix I -Current Induced Forces.}}

To verify the order-by-order energy conservation, the evaluation of
the work done by the electronic forces at a given order $n$ is necessary.
The work $W$ at a given order $n$ is
\begin{align*}
W^{\left(n\right)} & =\stackrel[0]{\tau}{\int}\left(\underset{\nu}{\sum}F_{\nu}^{\left(n\right)}\dot{X}_{\nu}\right)dt,
\end{align*}
where $\dot{X}_{\nu}$ is the slow time derivative of the classical
mechanical degree of freedom $\nu$, $F_{\nu}^{\left(n\right)}$ is
the $n$-th order of the adiabatic expansion of the electronic force
$F_{\nu}$. The expressions for $F_{\nu}^{\left(n\right)}$ up to
first-order are well known (see for example Refs. \onlinecite{bode2012Jan,deghi2021}).
They read
\begin{align*}
F_{\mu}^{\left(0\right)} & =-\frac{1}{2\pi i}\int\textrm{tr}\left\{ \varLambda_{\mu}G^{<}\right\} d\varepsilon,\\
F_{\mu}^{\left(1\right)} & =-\underset{\nu}{\sum}\gamma_{\mu\nu}\dot{X}_{\nu},
\end{align*}
where
\begin{align*}
\gamma_{\mu\nu} & =\frac{\hbar}{2\pi}\int\textrm{tr}\left\{ G^{<}\varLambda_{\mu}\partial_{\varepsilon}G^{R}\varLambda_{\nu}-G^{<}\varLambda_{\nu}\partial_{\varepsilon}G^{A}\varLambda_{\mu}\right\} d\varepsilon.
\end{align*}

\bibliography{Bibliography.bib}

\end{document}